\documentclass[twocolumn]{aastex63}

\pdfoutput=1

\usepackage{siunitx}
\usepackage[version=4]{mhchem}

\received{2020 March 4}
\revised{2020 March 29}
\accepted{2020 April 6}

\shortauthors{Sergeev et al.}

\begin{document}

\title{Atmospheric convection plays a key role in the climate of tidally-locked terrestrial exoplanets: insights from high-resolution simulations}

\correspondingauthor{Denis E. Sergeev}
\email{d.sergeev@exeter.ac.uk}

\author[0000-0001-8832-5288]{Denis E. Sergeev}
\affiliation{Department of mathematics\\
College of Engineering, Mathematics, and Physical Sciences, University of Exeter \\
Exeter, EX4 4QF, UK}

\author[0000-0002-4664-1327]{F. Hugo Lambert}
\affiliation{Department of mathematics\\
College of Engineering, Mathematics, and Physical Sciences, University of Exeter \\
Exeter, EX4 4QF, UK}

\author[0000-0001-6707-4563]{Nathan J. Mayne}
\affiliation{Department of astrophysics\\
College of Engineering, Mathematics, and Physical Sciences, University of Exeter \\
Exeter, EX4 4QL, UK}

\author[0000-0002-1485-4475]{Ian A. Boutle}
\affiliation{Department of astrophysics\\
College of Engineering, Mathematics, and Physical Sciences, University of Exeter \\
Exeter, EX4 4QL, UK}
\affiliation{Met Office, Fitzroy Road, Exeter, EX1 3PB, UK}

\author{James Manners}
\affiliation{Met Office, Fitzroy Road, Exeter, EX1 3PB, UK}
\affiliation{Global Systems Institute,\\
University of Exeter \\
Exeter, EX4 4QF, UK}

\author{Krisztian Kohary}
\affiliation{Department of astrophysics\\
College of Engineering, Mathematics, and Physical Sciences, University of Exeter \\
Exeter, EX4 4QL, UK}

\begin{abstract}
Using a 3D general circulation model (GCM), we investigate the sensitivity of the climate of tidally-locked Earth-like exoplanets, Trappist-1e and Proxima Centauri b, to the choice of a convection parameterization.
Compared to a mass-flux convection parameterization, a simplified convection adjustment parameterization leads to a \SI{>60}{\percent} decrease of the cloud albedo, increasing the mean day-side temperature by \SI{\approx 10}{\K}.
The representation of convection also affects the atmospheric conditions of the night side, via a change in planetary-scale wave patterns.
As a result, using the convection adjustment scheme makes the night-side cold traps warmer by \SIrange{17}{36}{\K} for the planets in our simulations.
The day-night thermal contrast is sensitive to the representation of convection in 3D GCM simulations, so caution should be taken when interpreting emission phase curves.
The choice of convection treatment, however, does not alter the simulated climate enough to result in a departure from habitable conditions, at least for the atmospheric composition and planetary parameters used in our study.
The near-surface conditions both in the Trappist-1e and Proxima b cases remain temperate, allowing for an active water cycle.

We further advance our analysis using high-resolution model experiments, in which atmospheric convection is simulated explicitly.
Our results suggest that in a hypothetical global convection-permitting simulation the surface temperature contrast would be higher than in the coarse-resolution simulations with parameterized convection.
In other words, models with parameterized convection may overestimate the inter-hemispheric heat redistribution efficiency.
\end{abstract}


\section{Introduction}
\label{sec:intro}
M-dwarf stars are the most prevalent stars in the galaxy and offer a higher chance for temperate terrestrial planets to be detected around them \citep{DressingCharbonneau2015}.
These stars are smaller and dimmer, compared to most other main sequence star types.
In order to receive a sufficient amount of stellar radiation and reside within the habitable zone (i.e. host liquid water on the surface), planets have to be in tight close-in orbits around M-dwarfs.
Proximity to their host stars is associated with strong tidal forces, meaning that the planet is likely to be in a ``tidally-locked'' synchronous rotation regime \citep{LeconteEtAl2015} and thus have a permanent day side and a permanent night side.
A temperate climate regime on a terrestrial tidally-locked planet can be destabilized in different ways and lose the ability to sustain liquid water.
On one hand, the atmosphere may transition into either a ``moist greenhouse'' or ``runaway greenhouse'' state, too hot for the water to be in a liquid state \citep{KastingEtAl1993,Goldblatt2015,WolfToon2015}.
On the other hand, the atmosphere may cool to the point of its gas constituents condensing to the surface and their greenhouse effect being removed, causing the temperatures plummet further.
Synchronously rotating planets with permanent cold regions (cold traps) on the night side can be particularly susceptible to atmospheric collapse, because atmospheric constituents accumulate here in permanent ice caps and cannot escape.
The moist/runaway greenhouse state and global glaciation demarcate the inner and the outer edges of the habitable zone, respectively \citep{KastingEtAl2014}.

For both the inner and outer boundaries of the habitable zone in the case of a tidally-locked planet, the effects of water vapor and clouds are of paramount importance \citep{LeconteEtAl2013,YangEtAl2013}.
At the inner boundary, convective clouds have a stabilizing effect on the climate: intense stellar irradiance of the dayside causes strong convection, which produces thick and deep convective clouds that raise the planetary albedo, thus reducing the amount of absorbed energy \citep{YangEtAl2013}.
The outer boundary of the habitable zone is controlled by the inter-hemispheric energy redistribution \citep{Wordsworth2015}.
In the absence of a global ocean or large geothermal flux, the atmospheric transport of energy is the only way of keeping the night side warm.
An essential part of this energy redistribution is horizontal latent heat transport, which depends on the intensity of the surface evaporation on the day side and water mixing ratio at the level of inter-hemispheric jets.
The water vapor and water clouds have the potential to control the greenhouse effect on the night side.
Accordingly, we can expect the day-side moist convection to be a key arbiter of planetary climates near both edges of the habitable zone.

Over the decades, the majority of planetary studies have employed relatively simple numerical codes, usually single-column models or low-resolution general circulation models (GCMs).
Individual convective cells are too small to be resolved by 3D GCMs (or not represented at all in the case of 1D models).
Running a global model with fully resolved convection is still extremely computationally demanding, especially over climate time scales.
While active model development is underway to allow for such simulations \citep[e.g. the LFRic project at the UK Met Office,][]{AdamsEtAl2019}, the overall effect of convection has to be \textit{parameterized} via physically-motivated but inevitably approximate schemes.
For Earth, convection is treated by parameterizations, which are ``tuned'' to the abundant observations available.
Even then, it remains one of the major causes of inter-model uncertainties regarding future climate projections \citep{SherwoodEtAl2014,CeppiEtAl2017}.
For exoplanets, no measurements of convective activity are available yet, so their atmospheres are usually modeled using simple approaches to convection, adopted from early generations of terrestrial GCMs (see Sec.~\ref{sec:conv_schemes}).
Many Earth-centric assumptions about convection are also applied to exoplanetary modeling, for instance the typical convection depth in the tropics \citep[e.g][]{YangAbbot2014}.

Recent research has started to explore the dependence of 3D GCM simulations of exoplanets to free parameters used in convection and cloud parameterizations.
\citet{YangEtAl2013} found that the stabilizing cloud feedback at the inner edge of the habitable zones is a robust phenomenon and the day-side cloud albedo is relatively insensitive to the choice of convection parameterization.
Using an improved version of the same GCM but in the context of the future Earth's climate, \citet{WolfToon2015} perturbed the microphysics and convection schemes and reported that the moist greenhouse state is relatively insensitive to such perturbations, because convection plays an increasingly minor role in the moist greenhouse conditions.
\citet{WayEtAl2018a} found that the dependence of the surface temperature of the Earth-like planet on the rotation rate is not substantially affected by the change in cloud condensate treatment in the convection parameterization. 
Confirming this finding, \citet{KomacekAbbot2019} noted that transition of the cloud cover for planets in various orbital regimes is overwhelmingly due to dynamics and should not be parameterization dependent.
It is worth noting, however, that the studies above employ only two GCMs and do not focus on the convection parameterization specifically.

\begin{deluxetable*}{llll}
\tablecaption{Planetary parameters used in this study \citep{GillonEtAl2017,Grimm2018,Anglada-Escude2016}.\label{tab:planets}}
\tablewidth{0pt}
\tablehead{
\colhead{Parameter} & \colhead{Units} & \colhead{Trappist-1e} & \colhead{Proxima b}
}
\startdata
Semi-major axis & AU & 0.02928 & 0.0485 \\
Orbital period ($2\pi \Omega^{-1}$) & Earth day & 6.10 & 11.186 \\
Obliquity & & \multicolumn{2}{c}{0} \\
Eccentricity & & \multicolumn{2}{c}{0} \\
Stellar irradiance ($S$) & \si{\watt\per\square\meter} & 900.482 & 881.700 \\
Planet radius ($r_p$) & \si{\kilo\meter} & 5804.071 & 7160.000 \\
Gravity & \si{\meter\per\second\squared} & 9.12 & 11.2 \\
\enddata
\end{deluxetable*}

There is clearly a need to explore convection on terrestrial exoplanets in a more comprehensive way, while also taking into account recent advancements in the theory of convection on Earth, especially the development of more fundamental and unified approaches \citep{RioEtAl2019}.
The lack of observational data for extraterrestrial convection can be circumvented by using a convection-permitting model that is able to simulate convective cells directly without reliance on a parameterization.
Currently, it is only feasible to use limited-area versions of these high-resolution models due to their computational cost \citep[e.g.][]{BrethertonKhairoutdinov2015,RioEtAl2019}.
Results from a regional convection-permitting model can then be used to benchmark and augment parameterizations in global GCMs, as has been done for Earth \citep[e.g.][]{MarshamEtAl2013,StrattonEtAl2018}.
Models that resolve convection have been used previously to improve our understanding of Solar System planets, namely Mars \citep[e.g.][]{SpigaEtAl2017} and Venus \citep[e.g.][]{LefevreEtAl2018}, but not exoplanets.
The exception is a short paper by \citet{ZhangEtAl2017}, who were the first to show the spatial variability of the atmosphere of a tidally-locked planet simulated with a kilometer-scale model.

Our study aims to show that the climate of terrestrial tidally-locked exoplanets is sensitive to the way convection is treated within a GCM.
Using the Met Office Unified Model, we first perform a series of coarse-resolution simulations with three different representations of convection to show that the circulation regime and the global cloud cover is substantially affected (see Sec.~\ref{sec:sens_conv}).
Importantly, the climate state of the whole planet and not just the substellar hemisphere (where convection occurs) is sensitive to the convection parameterization.
By modeling two different planets, Trappist-1e and Proxima Centauri b, we show that this effect is also planet-dependent.
We then use the same 3D GCM in a convection-permitting mode to obtain a fine-scale picture of atmospheric convection for a portion of the substellar hemisphere and to further explore the differences between the parameterized and explicit convection (see Sec.~\ref{sec:highres}).
Using an estimate of convection intensity on the day side, we hypothesize that a potential \textit{global} high-resolution simulation would enhance the day-night surface temperature contrast for a tidally-locked Earth-like planet (see Sec.~\ref{sec:dayside_impact}).

\section{Modeling Framework}
\label{sec:model}
We use the Met Office Unified Model (UM) in two set-ups: global coarse-resolution and regional high-resolution (explained in Sec.~\ref{sec:global_setup} and \ref{sec:regional_setup}, respectively).
The model equations of this 3D GCM are described in \citet{WoodEtAl2014} and represent the atmosphere as a deep non-hydrostatic fully compressible fluid.
Equations, solved using a semi-implicit semi-Lagrangian approach, are discretized on an Arakawa C-grid in the horizontal and a Charney-Phillips grid in the vertical.
Parameterized processes include subgrid-scale turbulence, convection (Sec.~\ref{sec:conv_schemes}), \ce{H2O} cloud and precipitation formation (with prognostic ice and liquid phases), and radiative transfer (handled by a correlated-$k$ scheme SOCRATES).
Full details of UM dynamics and physics can be found in \citet{WaltersEtAl2019} and references therein.
The UM has been adapted to simulate atmospheres of different planets, including gas giant planets \citep{MayneEtAl2014a,LinesEtAl2018b,MayneEtAl2019,DrummondEtAl2020} and rocky planets \citep{MayneEtAl2014b,BoutleEtAl2017,LewisEtAl2018}.

We use the orbital parameters of two confirmed terrestrial planets (see Table~\ref{tab:planets}): Trappist-1e and Proxima Centauri b (hereafter, Proxima b).
The stellar spectra of their corresponding host M dwarf stars are taken from BT-Settl \citep{RajpurohitEtAl2013}.
Both planets are assumed to be in 1:1 spin:orbit resonance, though this is of course not necessarily the case for a rocky planet closely orbiting an M-Dwarf star \citep{LeconteEtAl2015}.
Following \citet{BoutleEtAl2017}, all simulations have a \ce{N2}-dominated atmosphere with trace amounts of \ce{CO2} and \ce{H2O} amounting to a mean surface pressure of \SI{e5}{\pascal}.
We use the model in an aquaplanet regime, assuming a flat homogeneous surface at the lower boundary in a form of a slab water ocean \citep{Frierson2006}.
We choose the heat capacity of the slab layer to be \SI{e7}{\joule\per\kelvin\per\meter\squared}, corresponding to a depth of \SI{\approx2.4}{\meter}.
The surface albedo is fixed at $0.07$.

\begin{figure*}[!ht]
\begin{interactive}{js}{aas23243_fig1_interactive.zip}
\plotone{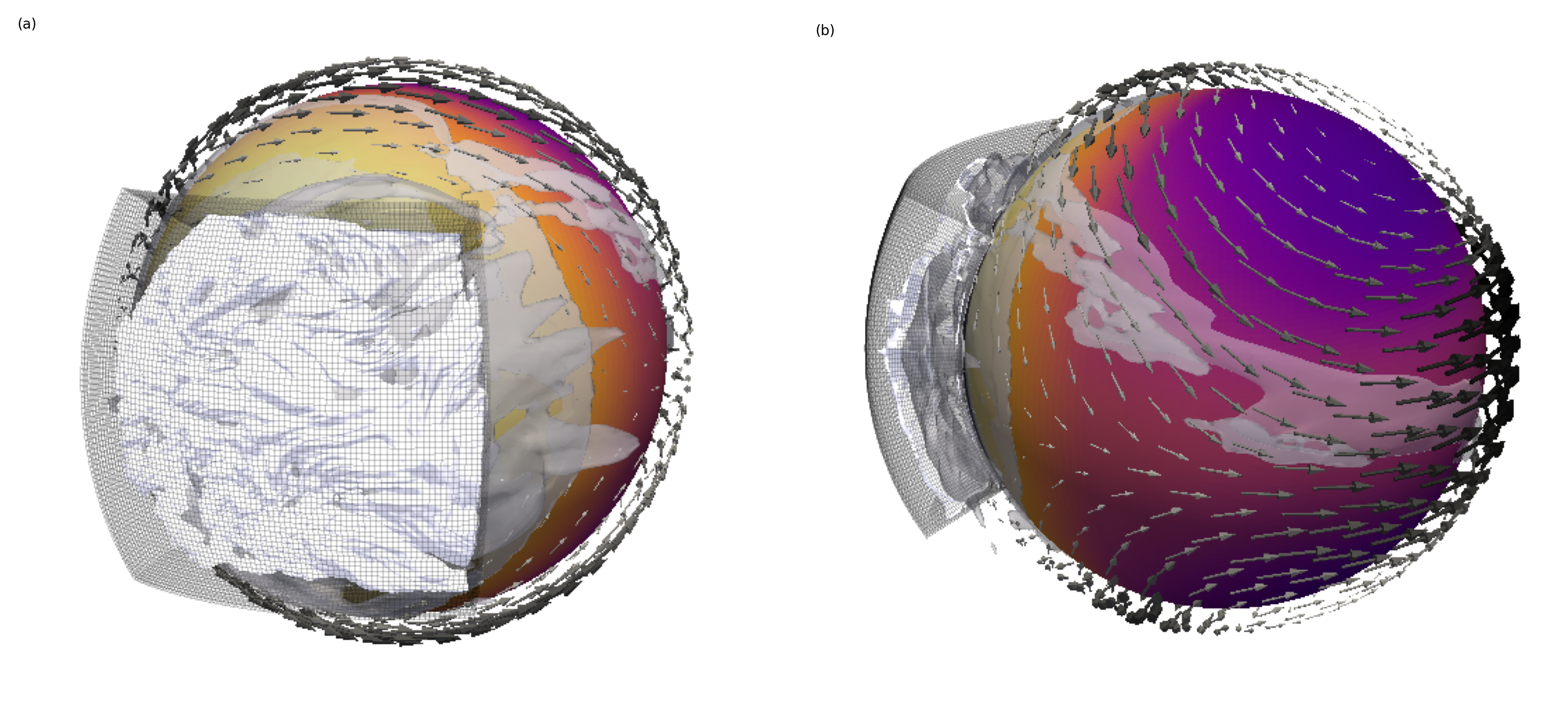}
\end{interactive}
\caption{Overview of the model set-up for the Trappist-1e case (see Tables ~\ref{tab:planets} and \ref{tab:runs}). The \emph{MassFlux} simulation covers the whole sphere, while the \emph{HighRes} simulation is shown by superimposing a high-resolution grid that covers only the substellar region. The graphic shows cloud condensate (white isosurfaces), surface temperature (shading), and free troposphere wind vectors (arrows), focusing on (a) the day side, with the \emph{HighRes} model domain and the cloud condensate isosurface, and (b) the night side of the planet. The cloud condensate is shown using a threshold of \SI{e-5}{\kg\per\kg} of total cloud condensate (liquid water plus ice). This figure is available online as an interactive zoom and rotatable figure.}\label{fig:overview3d}
\end{figure*}

\subsection{Global model}
\label{sec:global_setup}

The global set-up is equivalent to the Global Atmosphere 7 (GA7.0) configuration \citep{WaltersEtAl2019}.
Global simulations are performed with a horizontal grid spacing of \ang{2.5} in longitude and \ang{2} in latitude, with 38 vertical levels between the surface and the model top, located at \SI{\approx 40}{\km} height.
To account for unresolved convective processes, the global model uses a mass-flux convection parameterization, as described in more detail in Sec.~\ref{sec:conv_schemes}.
This control simulation is labeled \emph{MassFlux} (first row in Table~\ref{tab:runs}).
Next, the mass-flux parameterization is swapped for a simple convective adjustment scheme in the experiment labeled \emph{Adjust} (second row in Table~\ref{tab:runs}).
Finally, we switch off the convection parameterization entirely in the experiment labeled \emph{NoCnvPm} (third row in Table~\ref{tab:runs}).
The UM is integrated with a timestep of \SI{1200}{\s} until the steady thermal equilibrium regime is reached.
The results presented are averages over the last \SI{360}{\day} of \SI{1800}{\day} simulations (with Earth days as measure, $\SI{1}{\day} = \SI{86400}{\s}$).

\begin{deluxetable*}{lllll}
\tablecaption{Numerical simulations and their key differences. For more information on the convection parameterizations, see Sec.~\ref{sec:conv_schemes}.\label{tab:runs}}
\tablewidth{0pt}
\tablehead{
\colhead{Label} & \colhead{Model type} & \colhead{Grid spacing} & \colhead{Duration} & \colhead{Convection}
}
\startdata
\emph{MassFlux} & Global   & $\ang{2.5}\times\ang{2}$ & \SI{1800}{\day} & Mass-flux parameterization \\
\emph{Adjust}   & Global   & $\ang{2.5}\times\ang{2}$ & \SI{1800}{\day} & Adjustment parameterization \\
\emph{NoCnvPm}  & Global   & $\ang{2.5}\times\ang{2}$ & \SI{1800}{\day} & Explicit \\
\emph{HighRes}  & Regional & \SI{4}{\km}              & \SI{110}{\day}  & Explicit \\
\enddata
\tablecomments{\emph{MassFlux} also refers to the group of 46 additional simulations with perturbed mass-flux convection parameterization discussed in Sec.~\ref{sec:dayside_impact}.}
\end{deluxetable*}

\subsection{Regional high-resolution model}
\label{sec:regional_setup}
One of the strengths of the UM is the seamless approach to atmospheric modeling across spatial and temporal scales.
In practical terms, this means the same model core can be used for global climate prediction as well as kilometer-scale convection-permitting forecasts --- we label the latter \emph{HighRes}.
The main advantage of running a high-resolution model for studying convection is that it does not rely on a parameterization but simulates it explicitly (at least a large part of the convective spectrum).
Thus, in our high-resolution simulations (fourth row in Table~\ref{tab:runs}) the convection parameterization is switched off.
The rest of the physical parameterizations are set-up according to the Regional Atmosphere for Tropical Regions configuration \citep[RA1-T,][]{BushEtAl2019}.

We adopt a nested grid approach and perform high-resolution simulations in a limited-area domain.
The domain is a square with sides of \SI{\approx 6000}{\km} placed in the center part of the planet's dayside in order to capture the most vigorous convective regime (Fig.~\ref{fig:overview3d}a).
The horizontal grid spacing of the limited-area model is \SI{4}{\km} both for Trappist-1e and Proxima b and the number of vertical levels is 85.
A similar grid resolution is successfully used in weather simulations of the Earth atmosphere \citep[e.g.][]{MarshamEtAl2013,SergeevEtAl2017,StrattonEtAl2018}.
Here we show results of \SI{4}{\km} simulations, but we ran an additional simulation with a grid spacing of \SI{1.5}{\km}, which was qualitatively the same (not shown).
Sensitivity of the high-resolution simulations to the domain's location, size, and grid spacing is a fruitful avenue for future studies of tidally-locked exoplanets, but is beyond the scope of this paper.

Initial and boundary conditions are supplied by the global model with a time frequency of \SI{3600}{\s}, while the nested model time step is \SI{150}{\s}.
The run duration is 110 Earth days, consistent with the minimum run length recommendation of the Radiative-convective equilibrium model intercomparison project \citep[RCEMIP,][]{WingEtAl2018}.
An additional 365-day (one Earth year) experiment has not revealed a substantial model drift, i.e. the atmosphere is approximately in a steady-state when averaged over several planet orbits.
It should be also noted that the one-way nesting set-up used in our study has an important limitation, namely the absence of the feedback from the nested \emph{HighRes} model to the parent global model.
This will be discussed in more detail in Sec.~\ref{sec:dayside_impact}.

Our nested model set-up is somewhat similar to \citet{ZhangEtAl2017}, who applied the Weather and Research Forecasting (WRF) mesoscale model to explore dayside clouds at a high spatial resolution.
However, \citet{ZhangEtAl2017} used a completely different GCM for the parent and nested run.
Our modeling framework is self-consistent --- it uses the same dynamical core and physical parameterizations (except convection) in both set-ups.
This allows us to better separate the effects of the convection parameterization from other model components.

\subsection{Convection schemes in global simulations}
\label{sec:conv_schemes}
A standard convection parameterization representing the subgrid-scale energy transport associated with convection in coarse-resolution simulations at the UK Met Office is a mass-flux scheme based on \citet{GregoryRowntree1990}.
In addition, we use a simpler adjustment scheme developed by \citet{LambertEtAl2020}.

The \citet{GregoryRowntree1990} mass-flux scheme belongs to a class of the most advanced deterministic convection parameterizations.
It represents convection by decomposing the atmospheric column into different components: updrafts, downdrafts, and a subsiding environment \citep{Arakawa2004,RioEtAl2019}.
The collective effect of convective clouds is represented by a bulk vertical flux of heat, moisture, and momentum; as well as entrainment and detrainment of the ``background'' air in and and out of the convective plume, respectively.
The current mass-flux parameterization in the UM also includes downdrafts and convective momentum transport \citep{WaltersEtAl2019}.
Convection is triggered using an undilute parcel ascent and produces deep, mid-level, and shallow convection based on the convective available potential energy (CAPE) closure.
Similar parameterizations have been part of a few other terrestrial GCMs now applied to exoplanetary atmospheres \citep[e.g.][]{WolfToon2015,KopparapuEtAl2017,WayEtAl2017,Gomez-Leal2019}.

Applying a mass-flux scheme to exotic atmospheres of exoplanets has a few drawbacks, such as the complexity of the scheme and the fact that existing codes are developed for Earth meteorology.
A more transparent, albeit cruder, class of convection parameterizations are adjustment schemes \citep[e.g.][]{ManabeEtAl1965,Betts1986,Frierson2007}.
They are based on the concept of radiative-convective equilibrium and assume that the result of convection is the adjustment of the thermodynamic profile to a reference state.
In this study, we use the adjustment scheme developed by \citet{LambertEtAl2020}, which is similar to \citet{Betts1986}, but relies on different conditions for triggering convection --- a major uncertainty in parameterizing convection.
The main parameters of this scheme are the timescale of convection and critical relative humidity.
Having inherited a lot of parameterizations from the earliest generations of Earth models, the adjustment approach to the convection parameterization is also common in extraterrestrial GCMs \citep[e.g.][]{MitchellEtAl2009,WordsworthEtAl2011,LeconteEtAl2013,YangEtAl2013,TurbetEtAl2018,Koll2019,ThomsonVallis2019}.
We use the adjustment scheme in a set-up similar to that in the sensitivity experiments of \citet{WayEtAl2018a}, namely all cloud condensate in the convective updraft is forced to precipitate, implying that upper-level clouds can form only due to detrainment or large-scale upward motions.

Simulations without parameterized convection are also valuable and the parameterization is not required for numerical stability in modern GCMs.
Such experiments have been conducted for the Earth's climate and yielded results close to those with parameterized convection, at least in terms of the annual mean precipitation \citep{MaherEtAl2018}, so in the \emph{NoCnvPm} simulation we test what effect they would have for a tidally-locked terrestrial planet.

\begin{figure*}[!ht]
\includegraphics[width=\textwidth]{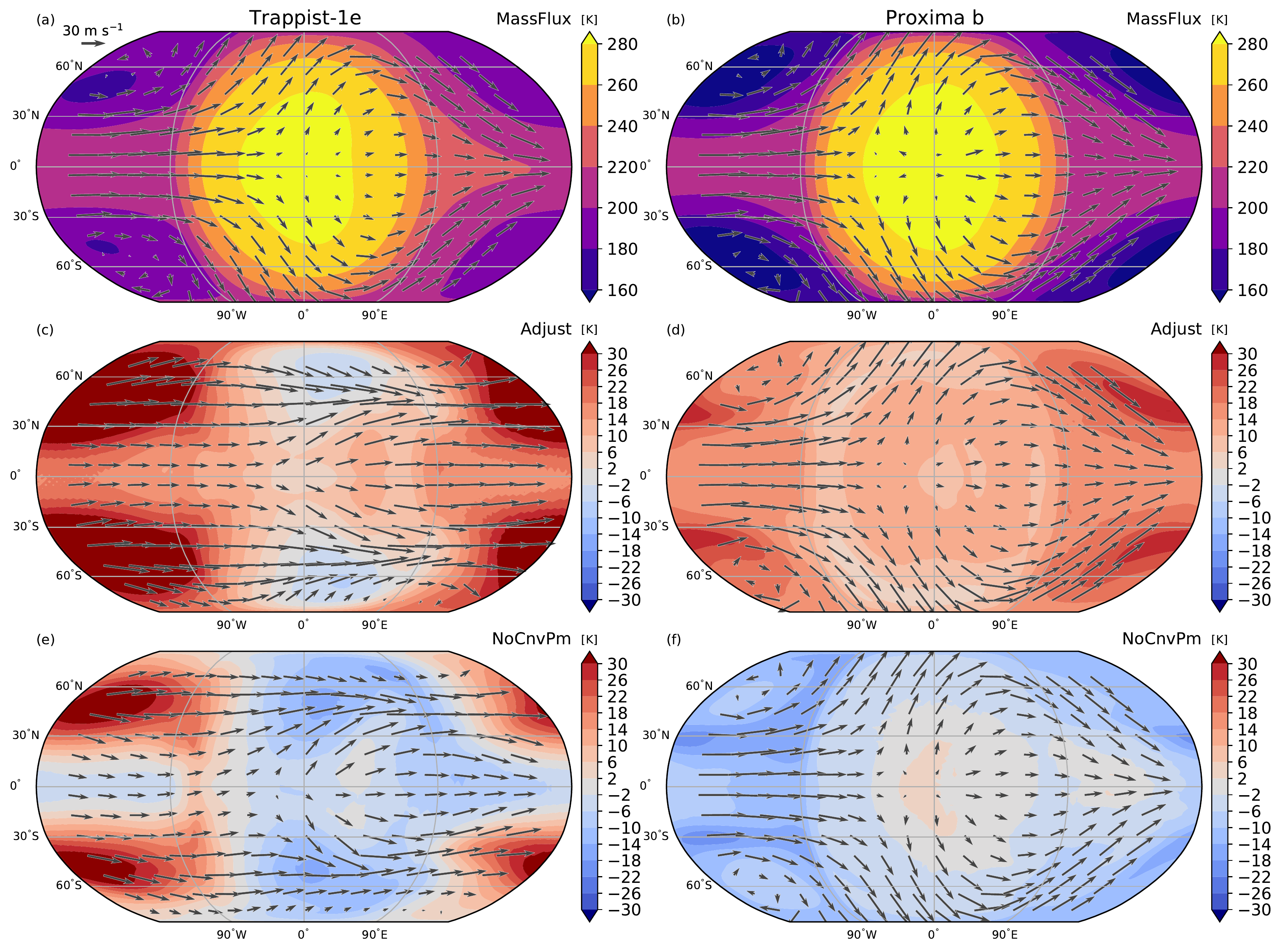}
\caption{Surface temperature (shading, \si{\kelvin}) and horizontal wind vectors in the upper troposphere (\SI{250}{\hecto\pascal} level) in the (left column) Trappist-1e case and (right column) Proxima b case in (a, b) \emph{MassFlux} and (c--f) sensitivity simulations. The surface temperature in the (c, d) \emph{Adjust} and (e, f) \emph{NoCnvPm} simulations is shown as the difference relative to the \emph{MassFlux} simulation. \label{fig:tsfc_winds}}
\end{figure*}

\section{Results}
We focus on the following major aspects of our numerical experiments.
In Sec.~\ref{sec:sens_conv} we present what the thermal structure (Sec.~\ref{sec:thermo}), circulation regime (Sec.~\ref{sec:circ}), and cloud distribution (Sec.~\ref{sec:cloud_precip}) of Trappist-1e and Proxima b are in the control (i.e. \emph{MassFlux}) simulation, and how sensitive they are to the choice of the convection parameterization (the \emph{Adjust} and \emph{NoCnvPm} simulations) in the UM.
How the convection parameterization affects the inter-hemispheric energy transport is analyzed in Sec.~\ref{sec:energy_transport}.
In Sec.~\ref{sec:highres}, we show novel results of high-resolution model experiments (\emph{HighRes}) and highlight differences between parameterized and explicit convection.
Finally, in Sec.~\ref{sec:dayside_impact} we estimate the planetary effect of substellar convection and, using a perturbed-parameter ensemble of global UM simulations, speculate what impact a hypothetical global convection-permitting simulation would make on the climate of tidally-locked planets.

\subsection{Mean global climate of Trappist-1e and Proxima b and its sensitivity to the convection parameterization}
\label{sec:sens_conv}

\subsubsection{Thermal structure}
\label{sec:thermo}

\begin{deluxetable*}{lcccccccccc}
\tablecaption{Mean global, mean day-side, mean night-side, maximum, and minimum surface temperature (\si{\kelvin}) in the control and sensitivity simulations of Trappist-1e and Proxima b.\label{tab:tsfc}}
\tablewidth{0pt}
\tablehead{
 & \multicolumn5c{Trappist-1e} & \multicolumn5c{Proxima b}\\
\colhead{Simulation} & $\overline{T_s}$ & $\overline{T_{s,d}}$ & $\overline{T_{s,n}}$ & $T_{s,max}$ & $T_{s,min}$ & $\overline{T_s}$ & $\overline{T_{s,d}}$ & $\overline{T_{s,n}}$ & $T_{s,max}$ & $T_{s,min}$
}
\startdata
MassFlux   &            231.8 &                260.0 &                202.7 &       288.3 &       177.1 &            226.6 &                261.7 &                190.5 &       287.3 &       148.7 \\
Adjust     &            249.3 &                268.3 &                229.3 &       294.3 &       214.1 &            241.0 &                272.0 &                209.0 &       299.3 &       166.0 \\
NoCnvPm    &            234.4 &                256.6 &                211.1 &       284.1 &       193.2 &            220.4 &                258.8 &                180.7 &       290.9 &       138.3 \\
\enddata
\end{deluxetable*}

In the control (\emph{MassFlux}) simulation of the Trappist-1e and Proxima b cases, the spatial distribution of the time-mean surface temperature ($T_s$) has a maximum at the substellar point and a minimum over the nightside mid-latitudes (Fig.~\ref{fig:tsfc_winds}a,b), typical for quickly-rotating tidally-locked planets with zero obliquity \citep[e.g.][]{HammondPierrehumbert2018,KomacekAbbot2019}.
In terms of absolute values, Proxima b has a larger $T_s$ contrast of \SI{138}{\K} compared to \SI{111}{\K} for Trappist-1e, due to much lower $T_s$ in the night-side cold traps: \SI{177}{\K} for Trappist-1e versus \SI{149}{\K} for Proxima b (Table~\ref{tab:tsfc}).
The day side, on the other hand, has a maximum of \SI{\approx 288}{\K} on both planets.
Previous 3D GCM studies of these planets report higher $T_s$, especially on the night-side of the planet \citep[e.g.][]{TurbetEtAl2016,Wolf2017,FauchezEtAl2020}.
As discussed by \citet{BoutleEtAl2017}, the differences stem from different representation of various parameterized processes in GCMs, such as the radiative effects of night-side clouds, subgrid-scale mixing in the boundary layer, and water vapor export from the day side.
The latter is investigated below in detail with respect to the day-side convection.

\begin{figure*}
\includegraphics[width=\textwidth]{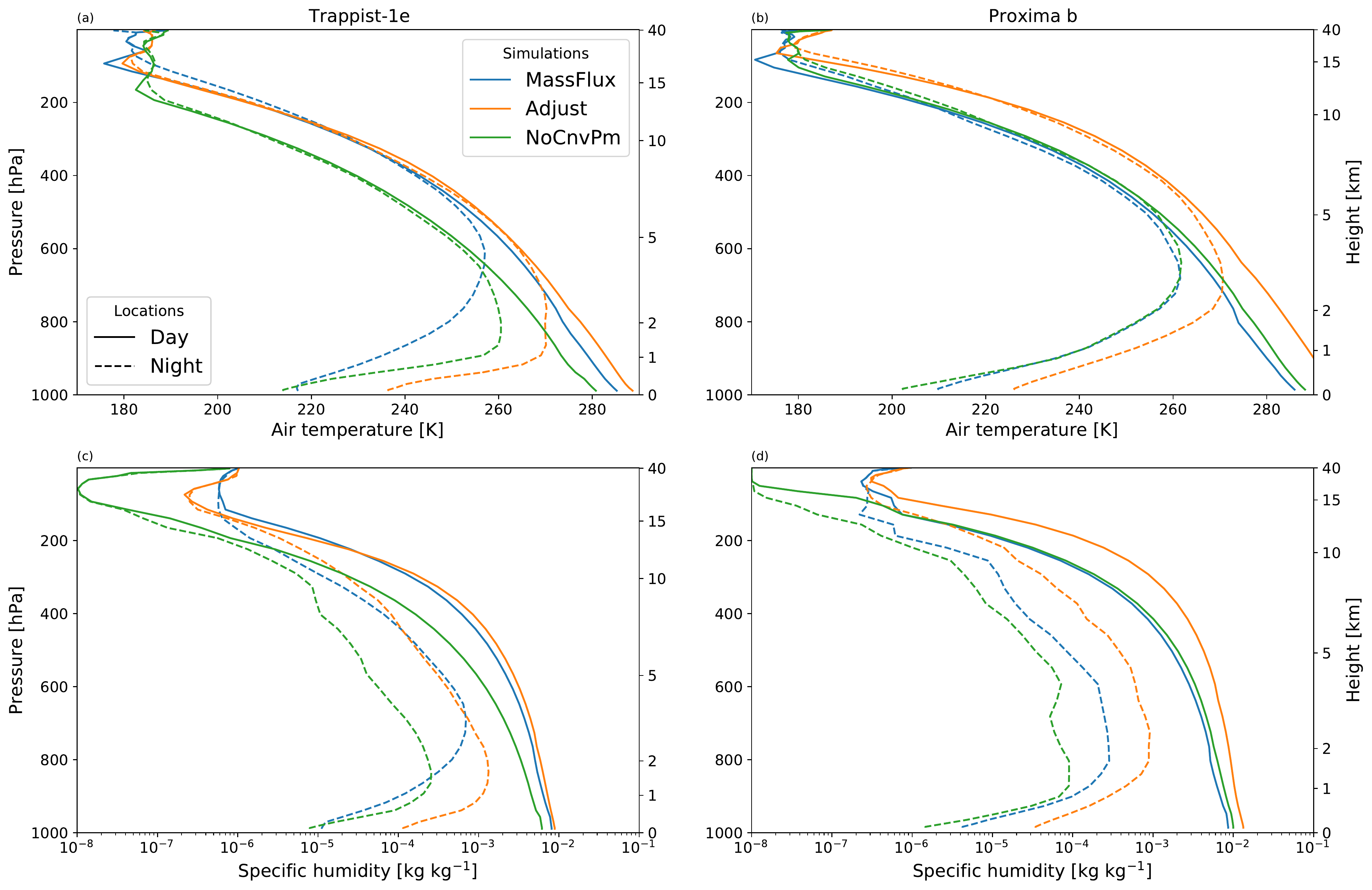}
\caption{Time average vertical profiles of (a, b) temperature ($T$, \si{\kelvin}) and (c, d) water vapor ($q$, \si{\kg\per\kg}) at the sub-stellar point (solid) and its antipode (dashed) in the (left column) Trappist-1e and (right column) Proxima b simulations with the (blue) \emph{MassFlux}, (orange) \emph{Adjust}, and (green) \emph{NoCnvPm} set-up. Note the logarithmic scale of the abscissa in (c) and (d).\label{fig:vprof_temp_shum}}
\end{figure*}

The day-side vertical stratification of the troposphere is defined by the irradiated surface beneath and emerging convective motions that redistribute the energy upward.
The temperature profile is quite steep near the surface and closely follows the dry adiabatic lapse rate (Fig.~\ref{fig:vprof_temp_shum}a,b).
Above the boundary layer (\SI{\approx 850}{\hecto\pascal}), the air temperature is closer to a moist adiabat, demonstrating the importance of moist convection, i.e. that the fall of temperature with height is offset by the release of latent heat of condensation.

The surface on the night side of both planets is extremely cold.
This is explained by both the absence of incoming solar radiation and small optical thickness because of low water vapor concentration.
In our simulations, the only source of energy for the night side is the atmospheric heat transport from the day side.
This idea is conceptually encapsulated in a two-column model of \citet{YangAbbot2014}, who note that the free troposphere should have a weak horizontal temperature gradient --- day-side and night-side vertical profiles in Fig.~\ref{fig:vprof_temp_shum}a,b indeed are the same above \SI{\approx 700}{\hecto\pascal}.
While the free tropospheric gradients are damped, the near surface air cools radiatively and then the boundary layer mixing cools the atmosphere above.
As a result, the night side is characterized by a very steep near-surface temperature inversion \citep[e.g.][]{Wordsworth2015,JoshiEtAl2020}.
In such a stably stratified atmosphere, vertical turbulent mixing of energy is suppressed, reinforcing the cold traps at the surface.
Note that the night-side surface temperature, through the day-night heat transport, is very sensitive to changes in the dayside surface temperature which is, in turn, set by a combination of processes.
In simulations with a dynamic ocean \citep[e.g.][]{DelGenioEtAl2018}, the response of the night-side temperature to the choice of the convection parameterization is likely to be different, but is out of scope of the present study.

Using the \emph{Adjust} set-up has broadly the same impact on both planets' mean $T_s$: it increases globally (\SI{18}{\kelvin} for Trappist-1e and \SI{14}{\kelvin} for Proxima b, see Table~\ref{tab:tsfc}).
As is visible in Fig.~\ref{fig:tsfc_winds}c and d, the greatest warming happens in the night-side cold traps, increasing $T_s$ to \SI{214}{\kelvin} for Trappist-1e and \SI{166}{\kelvin} for Proxima b.
On the day side, the change in $T_s$ is smaller --- only about \SI{10}{\kelvin} for both planets.
It is still positive everywhere for Proxima b (Fig.~\ref{fig:tsfc_winds}d), while for Trappist-1e it is positive in the tropics and negative in the mid-latitudes (Fig.~\ref{fig:tsfc_winds}c).
As a result, the mean temperature difference between the sub-stellar and the anti-stellar hemisphere ($\Delta T_{dn}$) is reduced from \SI{57}{\kelvin} to \SI{39}{\kelvin} for Trappist-1e and from \SI{71}{\kelvin} to \SI{63}{\kelvin} for Proxima b (Table~\ref{tab:tsfc}).

Temperature profiles in the \emph{Adjust} simulation largely follow those in the \emph{MassFlux} case, albeit with some deviations (Fig.~\ref{fig:vprof_temp_shum}, orange curves).
For Trappist-1e, using the adjustment scheme warms and moistens the low troposphere, but cools and dries the upper troposphere (Fig.~\ref{fig:vprof_temp_shum}a,c).
For Proxima b, it makes the air even warmer and moister throughout the entire troposphere  (Fig.~\ref{fig:vprof_temp_shum}b,d).
Altering the convection scheme in our model affects the temperature profile on the night side indirectly, via a change in the global wave pattern (see Sec.~\ref{sec:circ}).
In the \emph{Adjust} simulation of Trappist-1e (Fig.~\ref{fig:vprof_temp_shum}a), the night-side boundary layer is profoundly warmer, while in its lowest part the inversion is stronger, despite the surface also being warmer.

One unexpected result of our experiments is that changing the convection parameterization in the model alters the temperature not only on the day side, but also, and even more so, on the night side (where there is no convection).
In addition, the disparity between the global climate of Proxima b in our model and that simulated by \citet{TurbetEtAl2016} appears to be due to the difference in the convection scheme, as posited by \citet{BoutleEtAl2017}.
For example, the temperature profile in the \emph{Adjust} case (Fig.~\ref{fig:vprof_temp_shum}b) agrees better with the profile of \citet{TurbetEtAl2016}, who also used a convection adjustment scheme in their simulations.
The same argument can be made when comparing the night-side surface temperature minimum in the Trappist-1e case (Fig.~\ref{fig:tsfc_winds}) to that found using other GCMs \citep[e.g.][]{FauchezEtAl2020}.

\begin{figure*}
\includegraphics[width=\textwidth]{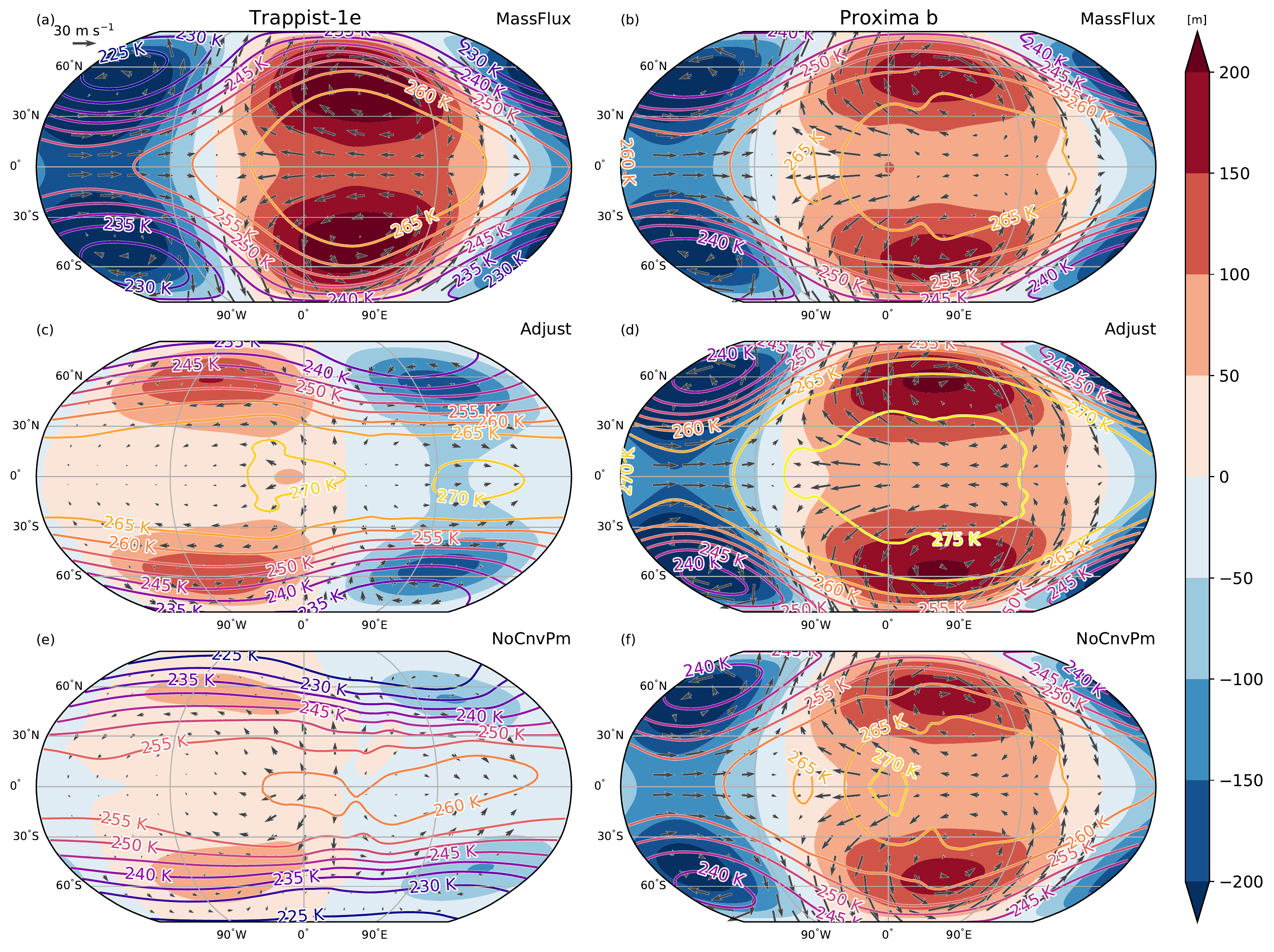}
\caption{Eddy geopotential height (shading, \si{\meter}) and eddy components of horizontal wind vectors \SI{250}{\hecto\pascal} and air temperature (contours, \si{\K}) at \SI{700}{\hecto\pascal} in the (left column) Trappist-1e case and (right column) Proxima b case in (a, b) \emph{MassFlux}, (c, d) \emph{Adjust}, and (e, f) \emph{NoCnvPm} simulations. Here the eddy component is taken as the deviation from the temporal and zonal mean.\label{fig:ghgt_winds}}
\end{figure*}

When we turn off the convection parameterization (the \emph{NoCnvPm} simulation) and let the UM remove thermal instability via the large-scale dynamics, the resulting global climate settles on a state different to that of the \emph{MassFlux} and \emph{Adjust} cases.
The globally averaged $T_s$ is about \SI{3}{\kelvin} higher for Trappist-1e, but \SI{6}{\kelvin} lower for Proxima b than in \emph{MassFlux} (Table~\ref{tab:tsfc}).
Higher $T_s$ of Trappist-1e is mostly due to the warming of the coldest night-side regions (Fig.~\ref{fig:tsfc_winds}e), similar to the \emph{Adjust} simulation.
Along the equator at almost all longitudes, but especially in the day-side mid-latitudes, $T_s$ is a few degrees lower compared to the control simulation.
In the \emph{NoCnvPm} simulation of Proxima b, surface cooling happens everywhere, except for the small region at the substellar point (Fig.~\ref{fig:tsfc_winds}f).
Meanwhile, its coldest regions become even colder: the temperature minimum drops to \SI{138}{\kelvin}

The temperature change in the substellar hemisphere suggests that in the \emph{NoCnvPm} simulation, the UM is less efficient in exporting energy from the hottest region of the planet.
One might indeed expect this outcome from a 3D Earth-like GCM.
However, in the Trappist-1e case the atmospheric circulation seems to mask this tendency by sufficiently warming the coldest regions, as discussed in Sec.~\ref{sec:circ}.

The atmospheric stratification at the substellar point and its antipode in \emph{NoCnvPm} is overall close to that in \emph{Adjust} for Trappist-1e and to that in \emph{MassFlux} for Proxima b (Fig.~\ref{fig:vprof_temp_shum}, green curves).
The \emph{NoCnvPm} simulation is noticeably colder and drier than \emph{MassFlux} for Trappist-1e, and especially dry in the upper atmosphere (Fig.~\ref{fig:vprof_temp_shum}c).
This cannot be explained by only invoking the Clausius-Clapeyron equation: while the temperature above \SI{200}{\hecto\pascal} is almost the same in all three sensitivity experiments, the water vapor content is lower by 2 or more orders of magnitude.
Instead, it is likely due to the less efficient moisture upwelling on the day side and thus weaker moisture transport between hemispheres.
Consequently, the magnitude of clear-sky longwave cooling on the night side is diminished (not shown).

\subsubsection{Circulation regime}
\label{sec:circ}

In all simulations, strong heating from the constantly irradiated surface on the day side results in a pressure minimum and a lower-level wind convergence.
This region is associated with strong upward motions that transport heat and moisture to the upper atmosphere, reducing convective instability.
In the upper troposphere above the sub-stellar point, the wind field is divergent (Fig.~\ref{fig:tsfc_winds}).
The main feature of the global wind field at this level is the eastward (prograde) jets, emerging in the atmosphere of many synchronously rotating planets \citep[e.g.][]{ShowmanPolvani2011,CaroneEtAl2015}.

The global circulation can be interpreted with the help of the eddy geopotential field (shown by shading in Fig.~\ref{fig:ghgt_winds}).
In the \emph{MassFlux} experiment, the quadrupole geopotential height distribution comprises two anticyclones to the east of the substellar point symmetric around the equator and two symmetric cyclones near the antistellar point (Fig.~\ref{fig:ghgt_winds}a,b).
Accordingly, the horizontal eddy wind field is characterized by alternating regions of divergence and convergence between the gyres in mid-latitudes.
This pattern corresponds to a particular wave solution in the shallow water equations, namely the westward propagating equatorial Rossby wave \citep[see Fig.~3e in ][]{KiladisEtAl2009}.
These planetary waves have a barotropic structure, as confirmed by analyzing the eddy geopotential field at different levels: cyclonic and anticyclonic gyres are aligned in height (not shown).
This is also indicated by the free-tropospheric temperature contours in Fig.~\ref{fig:ghgt_winds}a,b, largely aligning with the isolines of eddy geopotential.
 
The planetary-scale equatorial Rossby wave pattern dominates the atmospheric circulation both in the Trappist-1e and Proxima b \emph{MassFlux} cases.
However, there are subtle differences in their dynamical regimes, which can be expressed in terms of the ratio of the Rossby deformation radius and the Rhines length scale to the planet's radius \citep{Haqq-MisraEtAl2017}.
Following \citet{LeconteEtAl2013} and \citet{Haqq-MisraEtAl2017}, we estimate the non-dimensional Rossby radius of deformation, $\lambda_{Ro}$, and the non-dimensional Rhines length, $\lambda_{Rh}$, as
\begin{align}
    \lambda_{Ro} &= \frac{L_d}{r_p} = \sqrt{\frac{NH}{2\Omega r_p}},\\
    \lambda_{Rh} &= \frac{L_{Rh}}{r_p} = \pi \sqrt{\frac{U}{2\Omega r_p}},
\end{align}
where $L_d$ is the equatorial radius of deformation, $L_{Rh}$ is the Rhines length, $r_p$ is the planet radius, $\Omega$ is the planetary rotation rate, $N=\sqrt{g/\theta\partial\theta/\partial z}$ is the Brunt-V\"{a}is\"{a}l\"{a} frequency, $\theta$ is the potential temperature, $H=R_d T/g$ is the atmospheric scale height, $R_d$ is the dry air specific gas constant, $T$ is the air temperature, and $U$ is the zonal component of the wind vector.
Note the equatorial Rossby radius of deformation for continuously stratified fluids $L_d$ is used in the derivation above, and $\beta=2\Omega / r_p$.
Thermodynamic variables and the wind speed are calculated here as mass-weighted tropospheric averages (within \SIrange{0}{15}{\km} in height).

\begin{figure}
\includegraphics[width=0.5\textwidth]{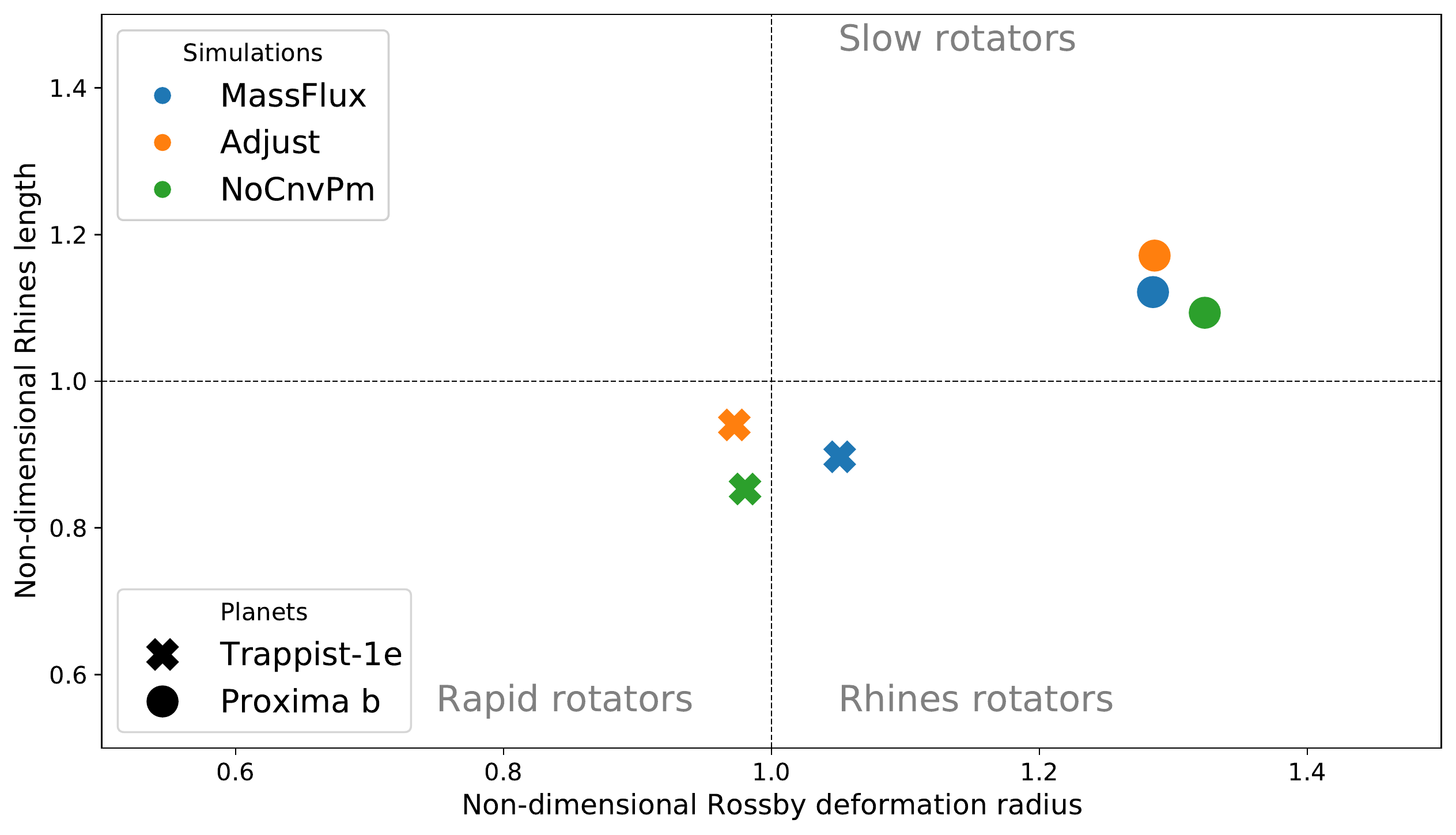}
\caption{Atmospheric circulation regimes of (crosses) Trappist-1e and (circles) Proxima b simulations with (blue) \emph{MassFlux}, (orange) \emph{Adjust}, and (green) \emph{NoCnvPm} set-up, defined by the estimates of the non-dimensional Rossby deformation radius ($L_d/R_p$, x-axis) and the non-dimensional Rhines length ($L_{R}/R_p$, y-axis).\label{fig:rossby_rhines}}
\end{figure}

Estimates of $\lambda_{Ro}$ and $\lambda_{Rh}$ reveal the planets' global circulation is close to the boundary between different rotation regimes (Fig.~\ref{fig:rossby_rhines}).
The atmosphere of Trappist-1e is in an especially delicate position, with the planetary Rossby wave length almost exactly matching the planet's radius ($\lambda_{Ro}\approx 1$).
For Proxima b, the Rossby deformation radius does not fit within the confines of the spherical geometry of the planet, as $\lambda_{Ro}\approx 1.25$.
Adding the Rhines length to the picture, we expect that zonal turbulence-driven jets can form and cause a departure from symmetry in the thermally direct circulation in the Trappist-1e case, because the length scale for the turbulent energy cascade is smaller, albeit not by much, than the planetary radius, i.e. $\lambda_{Rh}\lesssim 1$.
Proxima b is further away from the transitional boundary with $\lambda_{Rh} > 1$, indicating that its atmosphere is more dominated by thermally direct circulation \citep{Haqq-MisraEtAl2017}.

In the \emph{Adjust} simulation of Trappist-1e, the change in the atmospheric circulation regime explains the remarkable recession of cold spots on the night side.
To the first order, it is already hinted at by $\lambda_{Ro}$ crossing the boundary in Fig.~\ref{fig:rossby_rhines}, thus indicating that the circulation regime is closer to the rapid rotation than to the Rhines rotation \citep[as defined in][]{Haqq-MisraEtAl2017}.
The distribution of the eddy geopotential height is dominated by a pair of cyclonic gyres to the east of the substellar point and a pair of anticyclonic gyres to the west of it (Fig.~\ref{fig:ghgt_winds}c).
This pattern is almost a mirror-opposite of the distribution in the \emph{MassFlux} simulation (Fig.~\ref{fig:ghgt_winds}a), plus the gyres are shifted poleward and more zonally elongated.
Furthermore, the cyclone-anticyclone pairs have baroclinic structure, as illustrated by the large temperature gradient within their cores.
This structure corresponds to the extratropical Rossby wave pattern \citep[e.g.][]{CaroneEtAl2015}.

\begin{figure*}
\includegraphics[width=\textwidth]{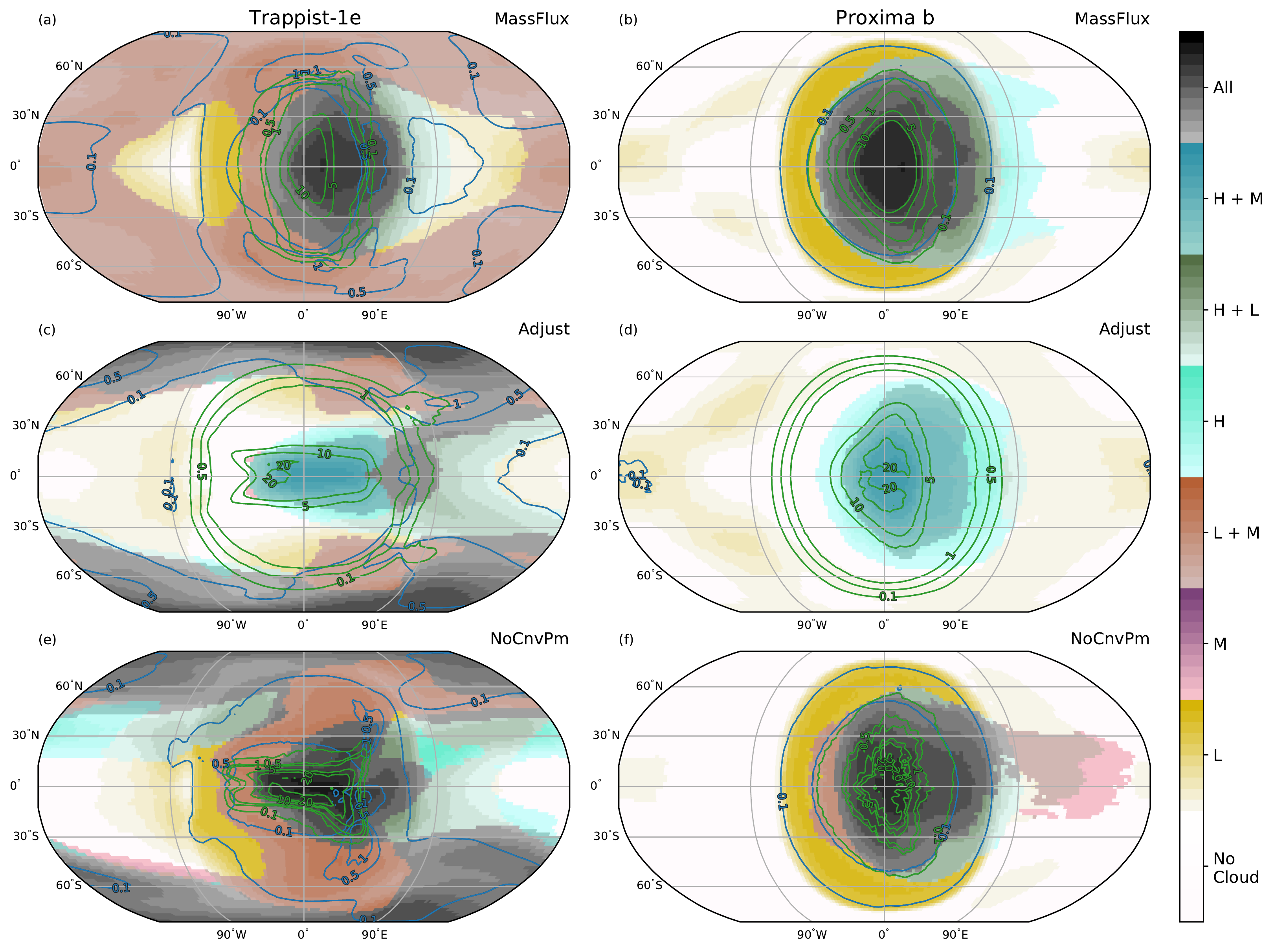}
\caption{Distribution of clouds and precipitation for the Trappist-1e and Proxima b cases in the control and sensitivity simulations. Low (L), mid-level (M), and high (H) cloud classes correspond to \SIrange{0}{2}{\km}, \SIrange{2}{5.5}{\km}, and \SI{>5.5}{\km} in altitude, respectively. The cloud cover is given by the maximum cloud fraction (0--1) on any model level. If only one type of cloud is present, the color scale shows that fraction, with a contour interval of $0.1$. If more than one type of cloud is present, the color scale shows the average of the two or three cloud types present. Note clouds with very low water content (optical depth $<0.01$) are filtered out. Mean precipitation (\si{\mm\per\hour}) in the form of rain and snow is shown by green and blue contours, respectively.\label{fig:global_cloud}}
\end{figure*}

Why is the atmospheric circulation in the Trappist-1e case defined by extratropical, rather than equatorial, Rossby waves?
We speculate that this is caused by the change in the zonal temperature gradient on the day side, which in turn is due to a different convection regime at the substellar point.
Comparing Fig.~\ref{fig:tsfc_winds}c to Fig.~\ref{fig:tsfc_winds}a, we see that in the \emph{Adjust} simulation the dayside, the tropics become warmer, while the substellar mid-latitudes become colder, thus enhancing baroclinicity towards the poles.
Larger baroclinicity enhances the extratropical Rossby waves, which then propagate meridionally and break in lower latitudes, decelerating the equatorial super-rotating jet \citep{MitchellVallis2010}.
Since the equatorial jet weakens, while the mid-latitude jets strengthen due to larger baroclinicity (via the thermal wind balance), the heat redistribution to the extratropics of the night-side becomes more effective, thus causing the cold traps to warm.

In \emph{NoCnvPm}, the mechanism of the night-side warming for Trappist-1e appears to be similar to that in \emph{Adjust}, confirmed by $\lambda_{Ro}<1$ (Fig.~\ref{fig:rossby_rhines}).
Even though the substellar point does not warm as in \emph{Adjust}, the day-side mid-latitude regions become substantially colder, so the baroclinicity increases, resulting in the extratropical Rossby wave-like response (Fig.~\ref{fig:ghgt_winds}e).

\begin{deluxetable*}{lcccccc}
\tablecaption{Global top-of-atmosphere shortwave ($CRE_{SW}$), longwave ($CRE_{LW}$), and total ($CRE$) cloud radiative effect (\si{\watt\per\square\metre}) in the control and sensitivity simulations of Trappist-1e and Proxima b.\label{tab:cre}}
\tablewidth{0pt}
\tablehead{
 & \multicolumn3c{Trappist-1e} & \multicolumn3c{Proxima b}\\
\colhead{Simulation} & $CRE_{SW}$ & $CRE_{LW}$ & $CRE$ & $CRE_{SW}$ & $CRE_{LW}$ & $CRE$
}
\startdata
\emph{MassFlux} &   -52.5 &   10.4 &   -42.1 & -58.3 &   15.7 &   -42.6 \\
\emph{Adjust} & -20.0 &    7.2 &   -12.8 & -20.4 &   10.2 &   -10.2 \\
\emph{NoCnvPm} & -57.4 &   13.1 &   -44.3 & -63.9 &   13.4 &   -50.5 \\
\enddata
\end{deluxetable*}

\subsubsection{Clouds and precipitation}
\label{sec:cloud_precip}

Due to deep convection, most of the day side of Trappist-1e and Proxima b is covered with multiple layers of clouds (shading in Fig.~\ref{fig:global_cloud}).
The water vapor carried upwards by convective updrafts eventually precipitates out, mostly in the form of rain (green contours in Fig.~\ref{fig:global_cloud}).
Snowfall forms a ring of weaker precipitation around the substellar region, coincident with lower temperatures (blue contours in Fig.~\ref{fig:global_cloud}).

While the day sides of both planets in the \emph{MassFlux} simulations are comparable in terms of the cloud cover and precipitation, the night sides are different.
Compared to the mostly cloudless Proxima b case, the night side of in the Trappist-1e case is almost completely covered by low and mid-level clouds and has more snowfall.
This disparity between the two simulated climates is associated with different energy balances: the longwave radiation cooling of Trappist-1e is partly due to the presence of clouds, while in the Proxima b case it is entirely due to clear-sky radiation (not shown).
Note the clouds also act as a greenhouse agent and emit radiation back to the surface.

In \emph{Adjust} (Fig.~\ref{fig:global_cloud}c,d), the day sides of both planets experience a severe reduction in low clouds with respect to \emph{MassFlux}: the substellar region is now shielded from the incessant stellar radiation only by mid- and high-level clouds.
For Trappist-1e, the high cloud layer is smaller and confined to the tropical latitudes, exposing large areas of the planet's surface to space.
For Proxima b, high-level clouds are broadly the same, but a large swath of day side is cloudless.
Consequently, the Bond albedo of both planets drops to $\approx 0.14$.

On the night side, using the \emph{Adjust} set-up results in different cloud response for Trappist-1e and Proxima b, as a consequence of the differences in the circulation regime.
The polar regions of Trappist-1e become completely enveloped with thick clouds in the \emph{Adjust} simulation, whereas the tropical regions are only covered with a thin veil of low clouds (and a high-level anvil at the eastern terminator, Fig.~\ref{fig:global_cloud}c).
The increase in extratropical clouds is associated with the increased role of baroclinic instabilities.
Meanwhile, the cloud layer on the night-side of Proxima b is only slightly thicker and does not extend higher than \SI{2}{\km} (Fig.~\ref{fig:global_cloud}d).
In the end, the changes in cloudiness and precipitation result in a slightly more positive night-side water balance (i.e. positive precipitation minus evaporation) on both planets compared to that of the \emph{MassFlux} experiment, with potential ramification for the global water supply being permanently trapped on the night side as ice over geologic timescales.

Due to the different treatment of condensed water in the convection parameterization (see Sec.~\ref{sec:conv_schemes}) and overall atmospheric warming, the total precipitation rate increases more than two-fold in the \emph{Adjust} experiments over the control.
The balance between the ``convective'' and the ``large-scale'' precipitation is tipped in favor of the former: from only \SIrange{20}{25}{\percent} of total precipitation being convective in \emph{MassFlux} to \SIrange{90}{95}{\percent} in \emph{Adjust}.
Note whilst this separation between convective and large-scale precipitation is artificial from the physical point of view, it is a necessary component of most GCMs.
More importantly, an incorrect balance between them can result in a different climate state as demonstrated here for tidally-locked Earth-like planets.

In \emph{NoCnvPm}, the day-side cloud pattern is close to that in \emph{MassFlux}, while the night-side pattern in the Trappist-1e case is close to that in \emph{Adjust}, in accordance with the global circulation response (see Sec.~\ref{sec:circ}).
These changes are accompanied by more concentrated precipitation over the substellar point.
As blue isolines in Fig.~\ref{fig:global_cloud}e,f demonstrate, the day-side rainfall field is much noisier, which is a manifestation of the grid-scale motions removing instability without the smoothing effect of the convection parameterization.
While the substellar region experiences a surge in precipitation, it declines significantly over the rest of the planet both in the Trappist-1e and Proxima b simulations, resulting in less water being deposited on the night-side.

\begin{figure*}
\includegraphics[width=\textwidth]{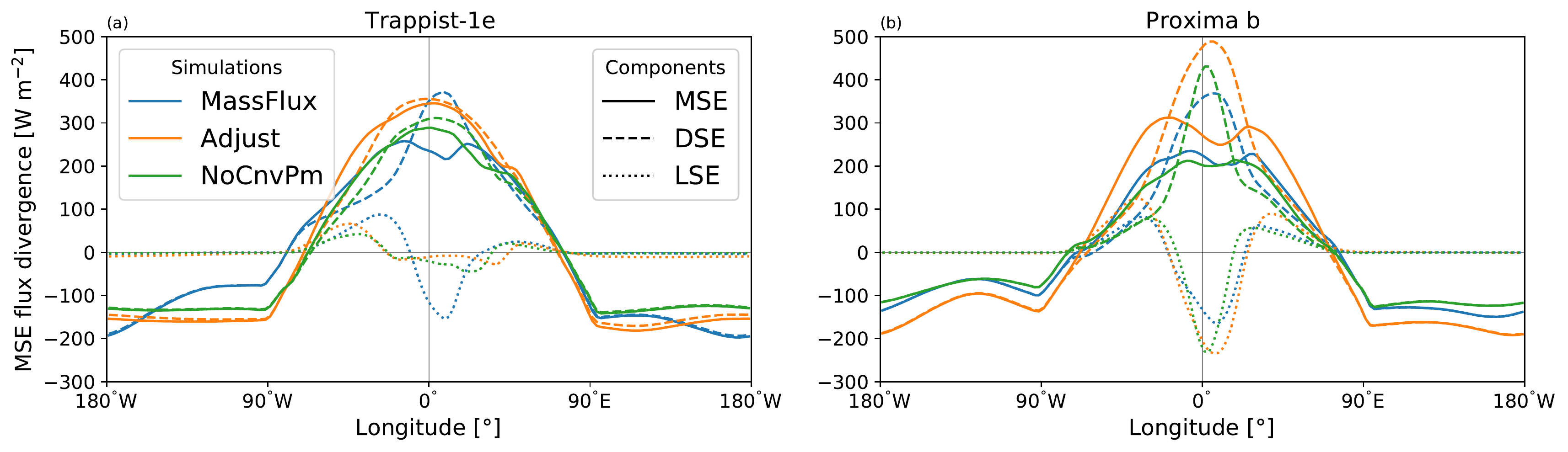}
\caption{Meridional average of vertically integrated divergence of (solid) MSE flux and its (dashed) dry and (dotted) latent parts ($\nabla\cdot\left(\vec u h \right)$, \si{\watt\per\square\meter}) in the (a) Trappist-1e case and (b) Proxima b case in (blue) \emph{MassFlux}, (orange) \emph{Adjust}, and (green) \emph{NoCnvPm} simulations.\label{fig:mse_div}}
\end{figure*}

Cloud distribution affects the fluxes of energy at the top of the atmosphere (TOA).
We use the TOA radiation fluxes to calculate the global cloud radiative effect ($CRE$) for each simulation (Table~\ref{tab:cre}).
$CRE$ is derived as the difference between the clear-sky and all-sky radiative flux \citep{CeppiEtAl2017}.
Day sides of tidally-locked terrestrial planets are mostly affected by its shortwave component ($CRE_{SW}$), which is negative, because \ce{H2O} clouds generally reflect solar radiation stronger than the planet's surface.
Both for Trappist-1e and Proxima b in the \emph{MassFlux} simulations, the magnitude of $CRE_{SW}$ is quite large (\SI{-52.5}{\watt\per\m\squared} and \SI{-58.3}{\watt\per\m\squared}, respectively) because of the high amount of thick clouds on the day side.
Clouds usually reduce outgoing terrestrial radiation, so the longwave $CRE$ ($CRE_{LW}$) is globally positive.
$CRE_{LW}$ is half as big for the planets in our study, compared to the present-day Earth.
To explain this, we recall that unlike its shortwave counterpart, $CRE_{LW}$ depends primarily on the emission temperature, which is a function of cloud altitude.
Whilst the night sides of the planets are characterized by the strong near-surface temperature inversion, the low-level and especially mid-level clouds are suspended in the warmest layer (Fig.~\ref{fig:vprof_temp_shum}a,b).
Consequently, the emission temperature of clouds is higher than that of the clear sky, resulting in a negative night-side (and lower globally) $CRE_{LW}$, notably in the case of Trappist-1e (see Fig.~\ref{fig:global_cloud}a and Table~\ref{tab:cre}).

In the \emph{Adjust} experiment, $CRE$ is dramatically reduced with respect to the control --- mostly due to the smaller magnitude of its shortwave component to \SI{\approx -20}{\watt\per\meter\squared} (Table~\ref{tab:cre}, middle row).
This is because the convection adjustment scheme used here immediately precipitates the available condensate, which results in fewer low- and mid-level clouds and thus lower $CRE_{SW}$.
Without the convection scheme (\emph{NoCnvPm}), the day side is enshrouded in thick clouds with a high-altitude anvil extending to the east (Fig.~\ref{fig:global_cloud}e,f), increasing the magnitude of $CRE_{SW}$ and making the total $CRE$ more negative (Table~\ref{tab:cre}, bottom row).
This effect is also evident in the increase of the day-side albedo.

\subsubsection{Energy redistribution to the night side}
\label{sec:energy_transport}

\begin{figure*}
\includegraphics[width=\textwidth]{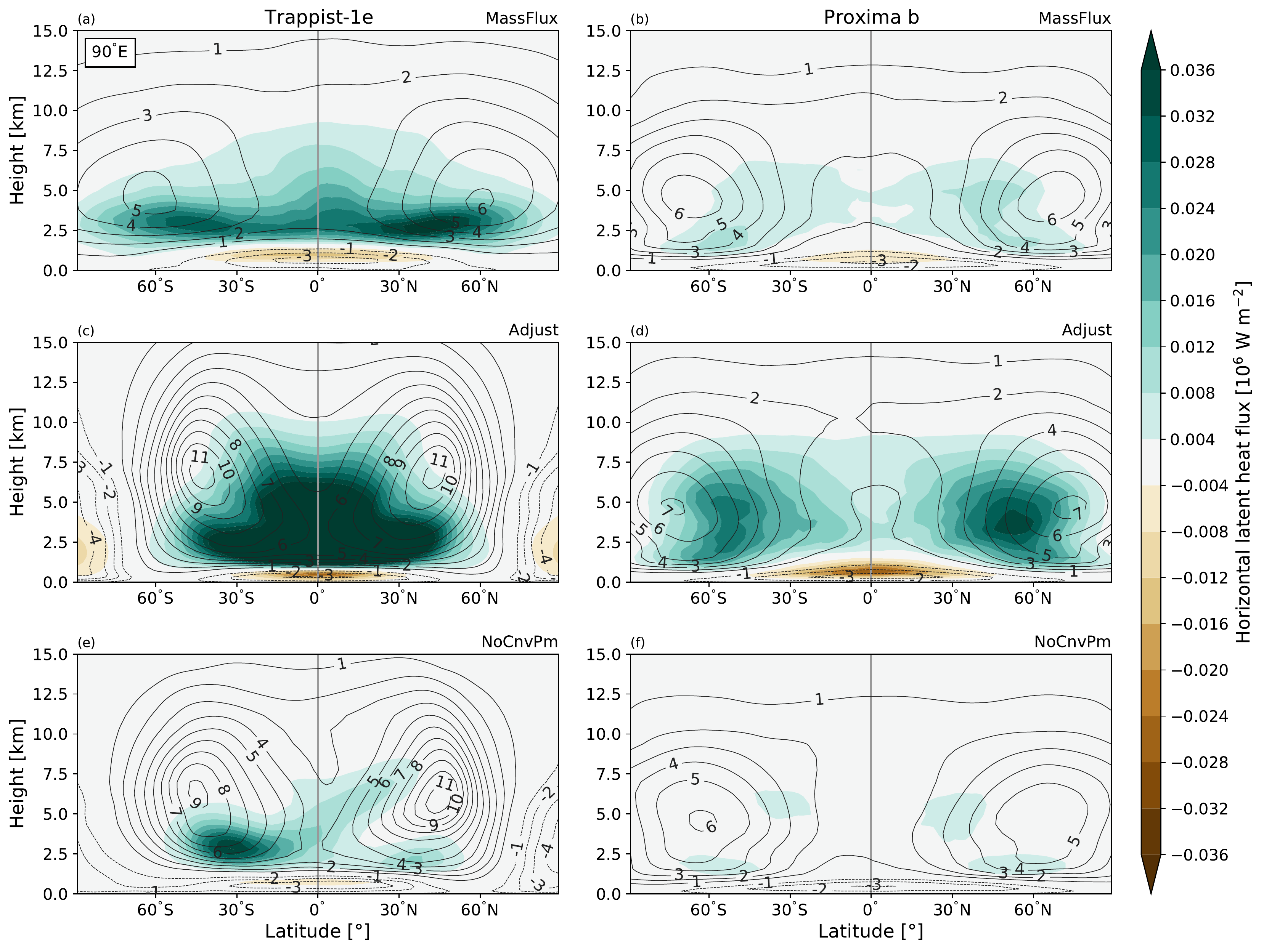}
\caption{Latitudinal cross-section of the zonal transport of sensible ($c_p T$, black contours, solid --- positive, dotted --- negative) and latent ($Lq$, color shading) heat (\SI{e6}{\watt\per\square\meter}) through the eastern terminator (\ang{90}E) in (left column) the Trappist-1e case and (right column) the Proxima b case in (a, b) \emph{MassFlux}, (c, d) \emph{Adjust}, and (e, f) \emph{NoCnvPm} simulations.\label{fig:vcross_heattrans}}
\end{figure*}

A useful estimate of the horizontal energy transport in planetary atmospheres is the mass-weighted vertical integral of the horizontal divergence of the moist static energy (MSE) flux, $\int_0^{z_{top}} \rho \nabla\cdot\left(\vec u h \right)dz$, where $h = c_p T + g z + L_v q$ and $\vec u$ is the horizontal wind vector.
Its average profile is presented in Fig.~\ref{fig:mse_div} as a function of longitude.
Principally, we see the MSE flux diverging in the substellar hemisphere and converging in the antistellar hemisphere in all our simulations.
The divergence of dry static energy (DSE, $c_p T + gz$) flux dominates over the latent component (LSE, $L_v q$), especially on the dry night side of the planet.
At the substellar point, the DSE flux divergence in the upper troposphere is substantially compensated by the convergence of LSE flux in the lower troposphere, reminiscent of the Hadley circulation on Earth.
A similar structure of the MSE transport divergence was also found in simulations of a slowly-rotating Earth-like aquaplanet \citep{MerlisSchneider2010}, albeit with higher amplitude due to the higher insolation received from the Sun.
Both in Fig.~\ref{fig:mse_div}a and b it is evident that the longitudinal profiles are not perfectly symmetric around the prime meridian and instead have peaks shifted east by a few degrees.
This is related to the eastward shift of the hottest part of the atmosphere due to the upper-level jets \citep[e.g.][]{ShowmanGuillot2002,PennVallis2018}; it also justifies the positioning of the high-resolution domain (see Sec.~\ref{sec:highres}).

To first order, profiles of MSE flux divergence are similar in the Trappist-1e case compared to those in the Proxima b case.
The largest difference is in the profiles of the latent component of MSE: the LSE flux convergence at the substellar point is much stronger in the Proxima b case (Fig.~\ref{fig:mse_div}b) in all three simulations, while in the Trappist-1e case it is quite weak in the \emph{MassFlux} simulation and almost negligible in the sensitivity simulations (Fig.~\ref{fig:mse_div}a).
At the same time, there is a larger contribution of the latent component to the MSE flux divergence on the night side of Trappist-1e, compared to that of Proxima b, which is in accord with the higher night-side column-integrated water vapor content in the Trappist-1e case (though the LSE component is still smaller than the DSE component).
In both cases, using the \emph{Adjust} set-up results in a more intense energy transport and in particular a larger share of the LSE flux divergence in the global energy transport.

To investigate the structure of this inter-hemispheric energy transport, we turn our attention to the zonal transport of sensible ($c_p T$) and latent ($L_v q$) heat through the plane of the eastern terminator (Fig.~\ref{fig:vcross_heattrans}), where the change in MSE flux divergence is the steepest (Fig.~\ref{fig:mse_div}).
The sensible heat transport clearly dominates the total heat transport, while its latent counterpart is $\approx 2$ orders of magnitude smaller.
Their structure in the vertical and horizontal is dictated by the wind field, but depends on the temperature and water vapor distribution too.
The day-to-night energy transport is the strongest at \SI{\approx 5}{\km} above the surface (\SI{\approx 500}{\hecto\pascal}), i.e. slightly lower than the jet stream maxima and coinciding with the altitude of the weak horizontal temperature gradient (Fig.~\ref{fig:vprof_temp_shum}a,b).
Close to the surface in the tropics, the heat transport is negative, partly canceling the net day-side divergence of the MSE flux (Fig.~\ref{fig:mse_div}).
The intensity of water vapor transport is largely correlated with the intensity of dry heat transport (cf. contours and shading in Fig.~\ref{fig:vcross_heattrans}).
Consequently, there are differences in the water vapor path on the night side, as mentioned above.

The heat flux cross-sections in Fig.~\ref{fig:vcross_heattrans}c--f elucidate the response to different convection parameterizations and how it differs between the two planets.
Consistent with the changes in the circulation regime, the dry heat transport maxima is almost doubled in the \emph{Adjust} experiment for Trappist-1e, while the water vapor transport substantially intensifies too and is more concentrated in the tropical region (Fig.~\ref{fig:vcross_heattrans}c).
With no convection parameterization (\emph{NoCnvPm}, Fig.~\ref{fig:vcross_heattrans}e), the dry heat transport is also greater than in the \emph{MassFlux} simulation, but the water vapor transport is substantially weaker.
Both in the \emph{Adjust} and \emph{NoCnvPm} cases, the day side to night side heat transport is closer to the equator, while the opposite flow emerges in the polar regions (Fig.~\ref{fig:vcross_heattrans}c,e).
The Proxima b case generally responds in a similar way, but the response is much more muted, especially in terms of the sensible heat transport cross-sections (Fig.~\ref{fig:vcross_heattrans}d,f).
Most importantly, the overall structure of the heat transport in the sensitivity experiments remains the same as in the \emph{MassFlux} experiment.

The global heat transport efficiency can be also expressed via the TOA energy fluxes.
Following \citet{LeconteEtAl2013}, we compare the ratio of the mean night-side TOA OLR over the mean day-side TOA OLR, referred to as the global heat redistribution parameter ($\eta$).
This parameter reaches its maximum when the horizontal transport is the largest, i.e. in the \emph{Adjust} experiments: $0.81$ for Trappist-1e and $0.67$ for Proxima b.
For Trappist-1e, $\eta$ is the lowest in \emph{MassFlux} ($0.70$) and for Proxima b in \emph{NoCnvPm} ($0.53$).
Thus selecting one convection parameterization or another for a terrestrial exoplanet can induce a difference of \SI{\approx 10}{\percent} in its thermal emission footprint.
Note the amount of water vapor on the night side of the planet affects the heat redistribution too by changing the emissivity of atmosphere \citep{LewisEtAl2018}.

\begin{figure*}
\includegraphics[width=\textwidth]{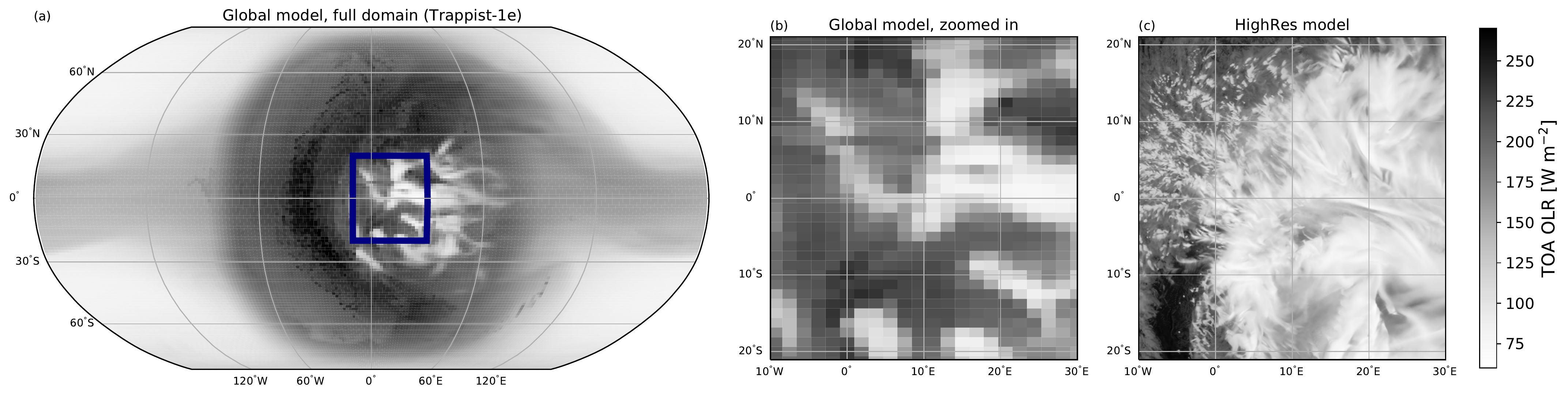}
\caption{Snapshot of top-of-atmosphere outgoing longwave radiation (\si{\watt\per\square\metre}) in the Trappist-1e simulation at \SI{102}{\day}: from the global model (a) over the whole planet, (b) zoomed in on the \emph{HighRes} domain; and (c) from the \emph{HighRes} simulation. The dark blue box in (a) shows the location of the \emph{HighRes} domain. \label{fig:trap1e_toa_olr_snap}}
\end{figure*}
\begin{figure*}
\includegraphics[width=\textwidth]{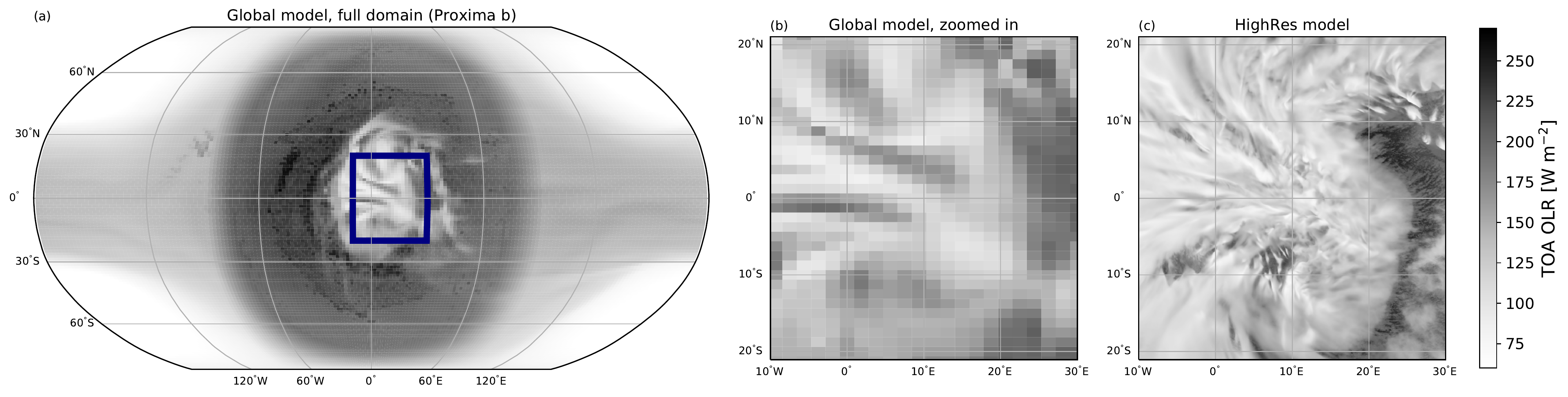}
\caption{Same as in Fig.~\ref{fig:trap1e_toa_olr_snap}, but for Proxima b at \SI{100}{\day}. \label{fig:proxb_toa_olr_snap}}
\end{figure*}
\begin{figure*}
\includegraphics[width=\textwidth]{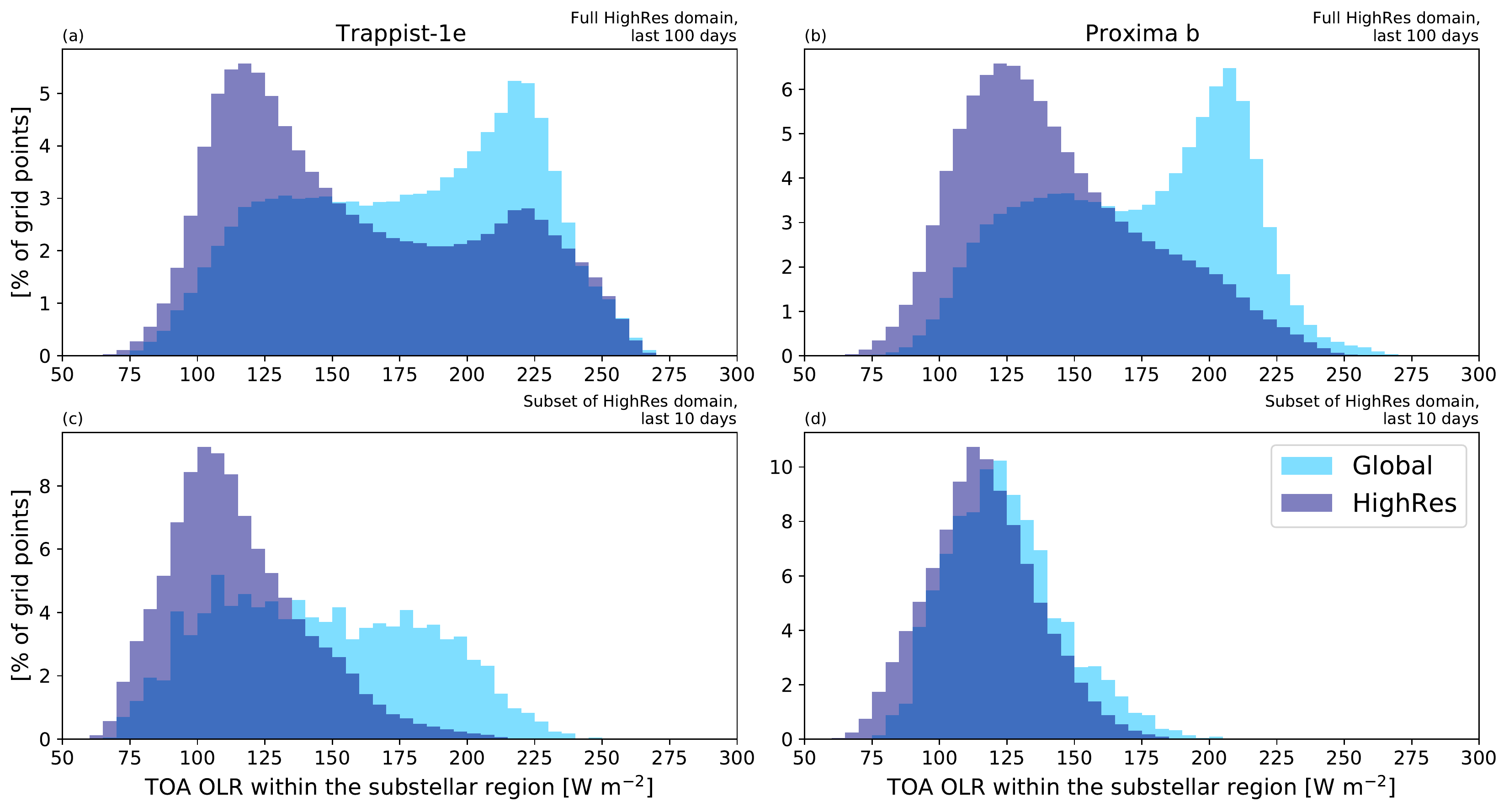}
\caption{Histograms of instantaneous TOA OLR values in (a, b) the full \emph{HighRes} domain, sampled every day over 100 days, and (c, d) in the central part of the \emph{HighRes} domain (shown by the blue box in Figs.~\ref{fig:trap1e_toa_olr_snap}, \ref{fig:proxb_toa_olr_snap}), sampled over 10 days, for (a, c) Trappist-1e and (b, d) Proxima b. Light blue bars show the global model values, while dark blue bars show \emph{HighRes} model values.\label{fig:toa_olr_hist}}
\end{figure*}

To sum up, the choice of convection parameterization is important for the global circulation and thermodynamic regime of tidally-locked planets.
By modulating the heat redistribution, they alter the thermal contrast between the day and night hemispheres.
The change in the thermal contrast is not likely to be observable by modern telescopes, implying that GCMs can rely on current convection parameterizations without affecting their synthetic phase curves too much.
Nevertheless, the change in the surface temperature, especially on the night side, is substantial (see Table~\ref{tab:tsfc}) and comparable to the impact of large variations in the incident stellar flux \citep[e.g.][]{KomacekAbbot2019}.
Furthermore, by looking at two slightly different planetary configurations, we show that the impact of convection on the global climate is sensitive to the planetary circulation regime in general and thus depends on external parameters such as the stellar irradiance, gravity, radius, and the rotation rate of the planet.
It should be noted that none of the existing parameterizations of convection is perfect, but most exoplanetary GCMs rely on them.
The results of this study demonstrate that caution should be taken when interpreting results of GCM simulations with one parameterization or another.
In the absence of observations, our best alternative to parameterized convection is a high-resolution convection-permitting model, used in the next section.

\subsection{High-resolution simulations}
\label{sec:highres}

The mean atmospheric state obtained in the global experiments described above is used to initialize a high-resolution limited-area simulation for each of the planets.
In both cases, initial and boundary conditions are supplied by the global model with the \emph{MassFlux} set-up.
We run the high-resolution experiment (\emph{HighRes}) for 110 Earth days (Sec.~\ref{sec:regional_setup}) and focus on the output over the final 10 simulation days.
The location of the nested domain is shown in Fig.~\ref{fig:overview3d}a, as well as in Fig.~\ref{fig:trap1e_toa_olr_snap}a and Fig.~\ref{fig:proxb_toa_olr_snap}a.
The domain is centered at (\ang{10}E, \ang{0}N) to capture the bulk of the convectively-active region.

\subsubsection{Horizontal inhomogeneity}
\label{sec:inhom}

Fig.~\ref{fig:trap1e_toa_olr_snap} and Fig.~\ref{fig:proxb_toa_olr_snap} present instantaneous TOA OLR in the (parent) global and \emph{HighRes} model simulations over the substellar region.
Besides reproducing individual convective plumes, shown by low values of OLR (i.e. high clouds), \emph{HighRes} also simulates cloud-free areas of subsidence more clearly than the coarse-resolution model (cf. middle and right panels in Fig.~\ref{fig:trap1e_toa_olr_snap} and Fig.~\ref{fig:proxb_toa_olr_snap}).
As a result, there is a clear separation between the two models in the TOA OLR histogram (Fig.~\ref{fig:toa_olr_hist}a,b).
In the global model, TOA OLR tends to be higher and has a peak at \SI{\approx 220}{\watt\per\square\meter}, corresponding to warmer and hence lower cloud tops (\SI{250}{\kelvin}, $\lesssim$\SI{5}{\km}).
The \emph{HighRes} histogram has only a minor secondary peak at this TOA OLR value (Fig.~\ref{fig:toa_olr_hist}a), while the absolute maximum is at \SI{\approx 120}{\watt\per\square\meter}, corresponding to colder and higher clouds (\SI{214}{\kelvin}, $\gtrsim$\SI{10}{\km}).
The decrease of OLR (\SIrange{\approx 15}{20}{\percent}) at the top of the atmosphere in \emph{HighRes} is compensated by the increase of the outgoing shortwave radiation.
In other words, the average Bond albedo of the substellar region increases by about \SI{3}{\percent} both for Trappist-1e and Proxima b.
It is important to note that the TOA OLR histograms are sensitive to the sampling region and time.
Using data from a central part of the domain and only over the last 10 days of the \emph{HighRes} simulation results in more similarity between the global and \emph{HighRes} data, especially for the Proxima b case (Fig.~\ref{fig:toa_olr_hist}d).
The remaining positive OLR bias of the global model in the Trappist-1e case (Fig.~\ref{fig:toa_olr_hist}c) can be attributed to the maximum in the liquid cloud condensate mixing ratio (see Sec.~\ref{sec:vert_struct}).
We aim to conduct more convection-permitting simulations with various grid configurations and nested domain placements in the future to explore the robustness of these results.

\begin{figure}
\includegraphics[width=0.45\textwidth]{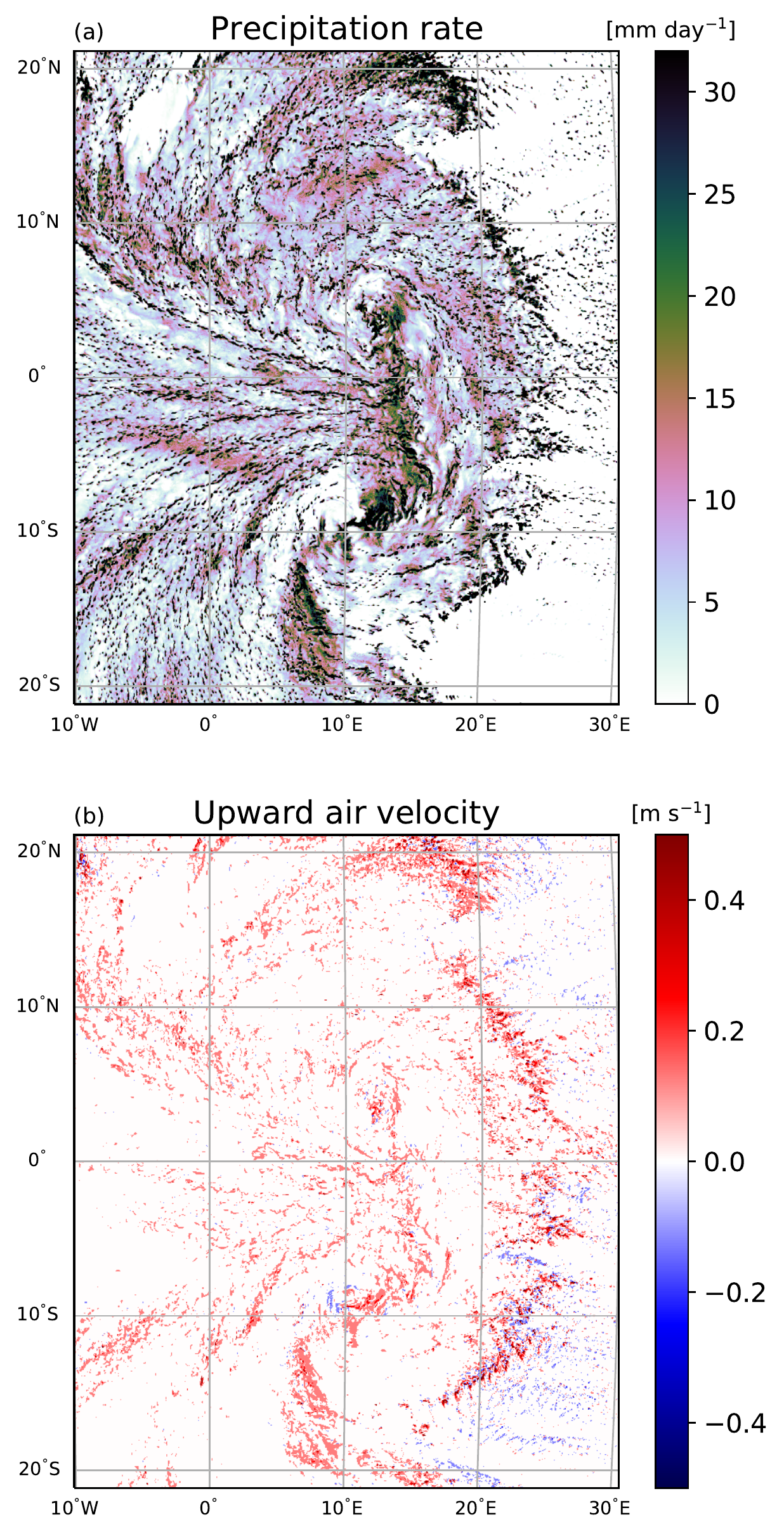}
\caption{Snapshot of convection in the substellar region in the Proxima b case at the end of the \emph{HighRes} simulation. Colors show (a) precipitation rate (\si{\mm\per\day}) and (b) upward vertical velocity (\si{\m\per\s}) at \SI{\approx 2500}{\m} altitude. \label{fig:proxb_precip_w}}
\end{figure}

Grid cells with clouds and intense precipitation (Fig.~\ref{fig:proxb_precip_w}a) correspond to the most intense convective updrafts, represented by positive vertical wind values, while dry patches are associated with downdrafts (Fig.~\ref{fig:proxb_precip_w}b).
In the global model, the vertical wind field is mostly positive, but has much lower absolute values in each grid cell.
Concomitantly, the precipitation rate in the global model is significantly weaker and spread over the whole region more evenly (not shown).
This results in the precipitation rate being capped at \SI{\approx 40}{\mm\per\day} in the global model, while the maximum precipitation rate in the \emph{HighRes} model exceeds \SI{100}{\mm\per\day} (Fig.~\ref{fig:proxb_precip_w}a).
There are \SI{>40}{\percent} more cells with low precipitation rate (\SI{\leq 1}{\mm\per\day}) in the \emph{HighRes} model than in the global simulation, reminiscent of the change in rainfall on Earth with and without parameterized convection \citep{MaherEtAl2018}.
Qualitatively speaking, the \emph{HighRes} simulation is able to reproduce a wider spectrum of atmospheric circulations, including individual convective cells, mesoscale frontal structures, and cyclonic vortices.

\subsubsection{Vertical cloud structure}
\label{sec:vert_struct}
\begin{figure*}
\includegraphics[width=\textwidth]{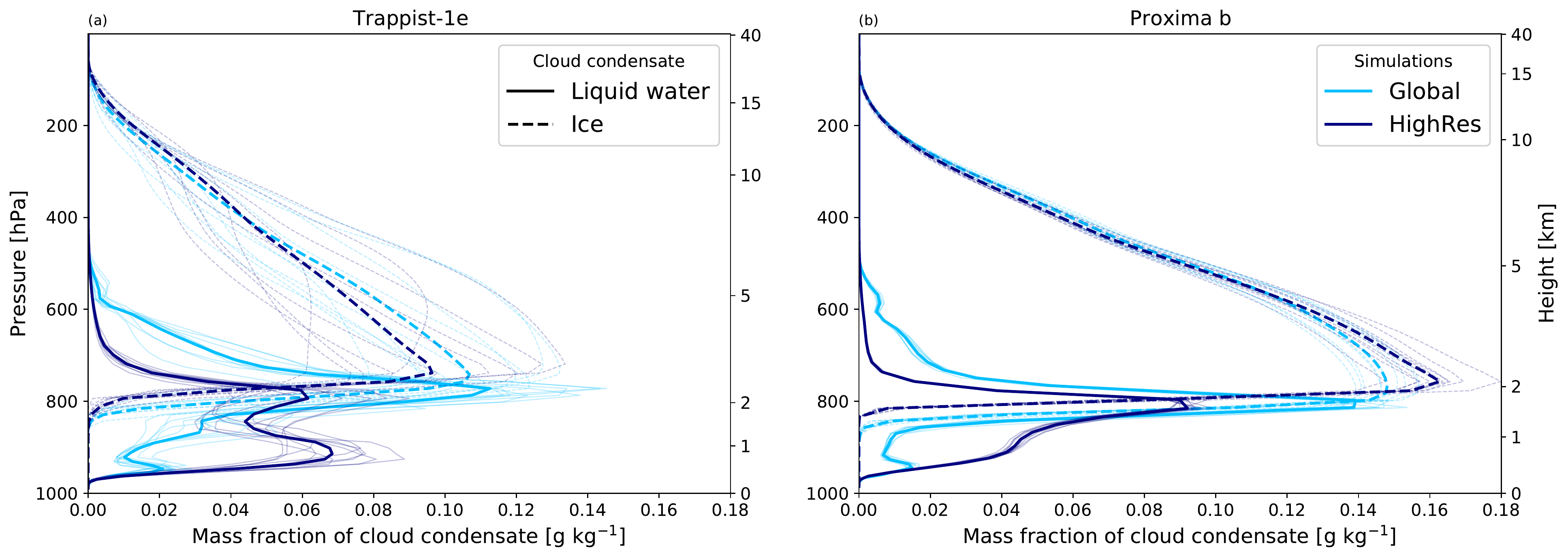}
\caption{Vertical profiles of the cloud (solid) liquid water and (dashed) ice mixing ratio averaged over the center part of the \emph{HighRes} domain in (light blue) global and (dark blue) \emph{HighRes} simulations of (a) Trappist-1e and (b) Proxima b. Thin lines represent spatial averages for each day of the 10-day analysis window, thick lines are the 10-day mean.\label{fig:cloud_vprof_lam}}
\end{figure*}

\begin{figure*}
\includegraphics[width=\textwidth]{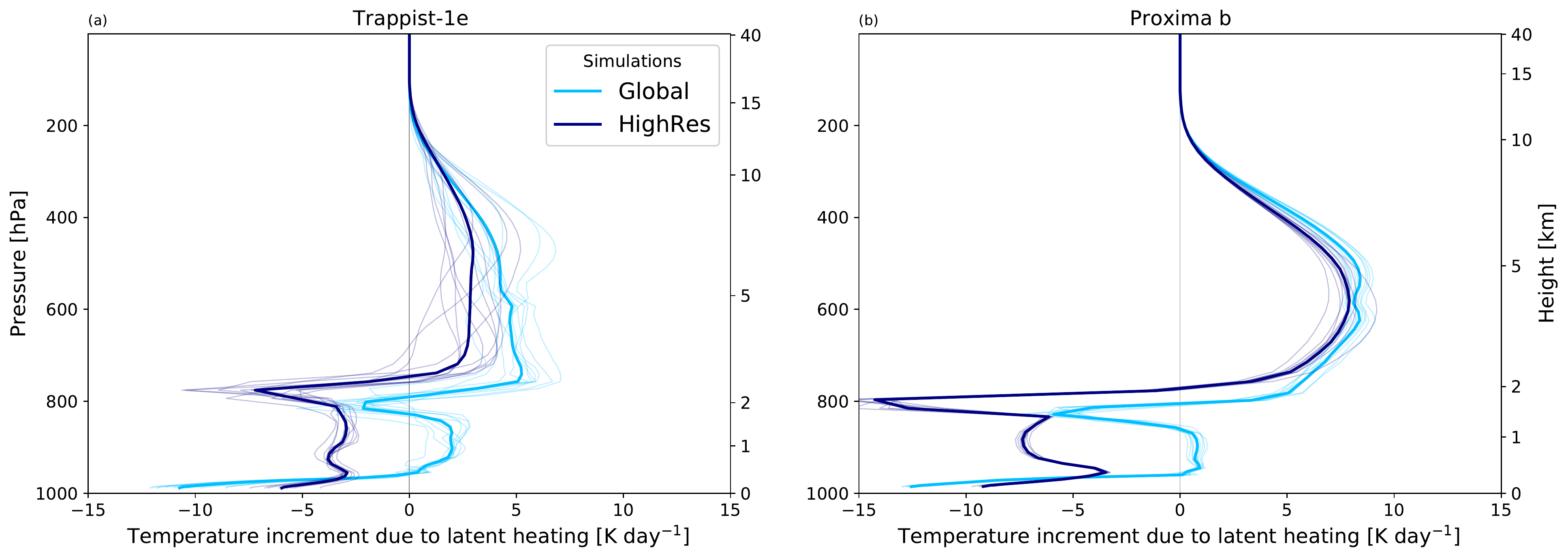}
\caption{As in Fig.~\ref{fig:cloud_vprof_lam}, but for temperature increments due to latent heating.\label{fig:t_incr_vprof_lam}}
\end{figure*}

Despite the spatial variability of convective activity and TOA energy fluxes, the average atmospheric stratification in the substellar region is rather similar between the global simulation and the \emph{HighRes} simulation.
The temperature and water vapor in the \emph{HighRes} simulation essentially follow the \emph{MassFlux} lapse rates shown in Fig.~\ref{fig:vprof_temp_shum}, though there is some minor deviation: the lower troposphere is slightly colder and drier in the Trappist-1e case and slightly warmer and moister in the Proxima b case --- resembling the effect of the \emph{NoCnvPm} simulation.
Note the vertical profiles represent 10-day averages, and instantaneous profiles differ more.

More prominent differences between the global and \emph{HighRes} experiments appear in the vertical profiles of cloud liquid and frozen water content (Fig.~\ref{fig:cloud_vprof_lam}).
In general, these profiles show a transition between liquid-phase clouds at lower model levels, to mixed-phase clouds within \SIrange{\approx 1.7}{3.5}{\km}, to ice-phase clouds aloft.
Parameterized convection in the global simulation, interacting with the cloud scheme, tends to underproduce clouds in the boundary layer (2--3 times less liquid water) when compared to the \emph{HighRes} model.
However, in the mixed-phase cloud layer above, the liquid water content is overestimated by the global model, resulting in an overall higher liquid water content within the substellar region.
The concentration of ice crystals in the global model, on the other hand, has a smaller bias relative to the \emph{HighRes} model, though it is positive in the Trappist-1e case and negative in the Proxima b case.
With regards to the mixed phase \citep[where ice crystals comprise between 20 to 80\% of the total cloud condensate,][]{SergeevEtAl2017}, the global simulation makes this type of cloud thicker by several hundred meters than that in \emph{HighRes}.
This has an impact on the formation of precipitation, as well as graupel-type hydrometeors and hence the generation of lightning \citep{TostEtAl2007}.
The latter may then alter the atmospheric composition, potentially detectable by future telescopes \citep{ArdasevaEtAl2017}.

The TOA energy balance over the substellar point is predominantly affected by the upper layers of convective clouds, therefore the disparity in frozen cloud condensate between coarse- and high-resolution simulations could be crucial.
As Fig.~\ref{fig:cloud_vprof_lam} shows, the cloud ice profiles at the very top of the domain are congruent in both models, which is an unexpected result given the noticeable differences in TOA fluxes mentioned above.
A deeper investigation of the model's microphysics scheme and its interaction with the convection parameterization is required to reveal the answer to this.

Within the lower-to-mid troposphere, radiation heating is much weaker than the temperature changes due to the latent heating or cooling (not shown).
Differences in cloud structure between the global and \emph{HighRes} model affect this heating profile.
The \emph{HighRes} model, having more low-level liquid-phase clouds (Fig.~\ref{fig:cloud_vprof_lam}), tends to have a strong negative latent cooling due to the evaporation of cloud condensate (Fig.~\ref{fig:t_incr_vprof_lam}).
In the global simulation, on the other hand, the cloud-related temperature change is smaller, hovering around zero at these vertical levels.
At approximately \SI{800}{\hecto\pascal} (\SI{\approx 2}{\km}) we see a local cooling peak, which is especially prominent in the Proxima b case (Fig.~\ref{fig:t_incr_vprof_lam}b).
Above this level, the latent heating starts to dominate both in the global and \emph{HighRes} simulations, with shortwave heating overtaking in the upper troposphere (not shown).
The rate of the latent heat release is stronger in the global model, though the difference is obscured by internal variability (see e.g. Fig.~\ref{fig:t_incr_vprof_lam}a).
Nevertheless this contrast in heating rates results in the different upper-level flow divergence, as we discuss in the next section (Sec.~\ref{sec:dayside_impact}).

\subsection{Global impact of resolved convection}
\label{sec:dayside_impact}

The benefits of running a high-resolution convection-permitting model do not cease at reproducing the structure of mesoscale weather systems and fine-scale features of the cloud cover.
It also improves the energy transport in the substellar region.

In the previous section (Sec.~\ref{sec:highres}) we demonstrate that the global coarse-grid model performs reasonably well when compared to the \emph{HighRes} model in terms of the mean state of the convective region.
This is perhaps unsurprising given the level of sophistication of the mass-flux parameterization used in the global model, as well as the Earth-like atmospheric temperature and pressure used in this study.
However, as indicated by Fig.~\ref{fig:vprof_temp_shum}, having relatively similar thermodynamic profiles in the substellar region can still result in different climate conditions globally (\emph{MassFlux} versus \emph{Adjust} or \emph{NoCnvPm}).
Moreover, we have shown that the vertical distribution of latent heating is positively biased in the global model (Fig.~\ref{fig:t_incr_vprof_lam}), prompting us to estimate the potential global impact of the \emph{HighRes} simulation.

Our current model set-up does not allow us to estimate this impact directly due to the one-way grid nested configuration.
Nevertheless, we can indicate a likely trend by identifying a statistical relationship between the metric of convective activity within the substellar region and the state of the resultant climate in global simulations.
For the latter we use the mean surface temperature difference between the day side and the night side of the planet ($\overline{\Delta T_{dn}}$).
Substellar convective activity is expressed via the horizontal flow divergence in the free troposphere within the domain of the \emph{HighRes} experiment ($\overline{\nabla\cdot\vec{u}_{ss}}$).
Both parameters can be calculated for global simulations, while only $\overline{\nabla\cdot\vec{u}_{ss}}$ is available in the \emph{HighRes} simulation (because it is a regional simulation).
We thus aim to create a regression using a series of global simulations and then suggest $\overline{\Delta T_{dn}}$ from a given value of the flow divergence in \emph{HighRes}.
To build the regression we conduct an additional series of global experiments with a suite of slightly different configurations of the mass-flux convective scheme.
The scheme parameters are perturbed one at a time, while the rest are the same as in the control.
This \emph{MassFlux} ``ensemble'' comprises 47 realizations, including the control, which has been analyzed above in more detail (Sec.~\ref{sec:sens_conv}).
The list of perturbed parameters with corresponding values is given in the appendix (Table~\ref{tab:grcs_ens}).

\begin{figure*}
\includegraphics[width=\textwidth]{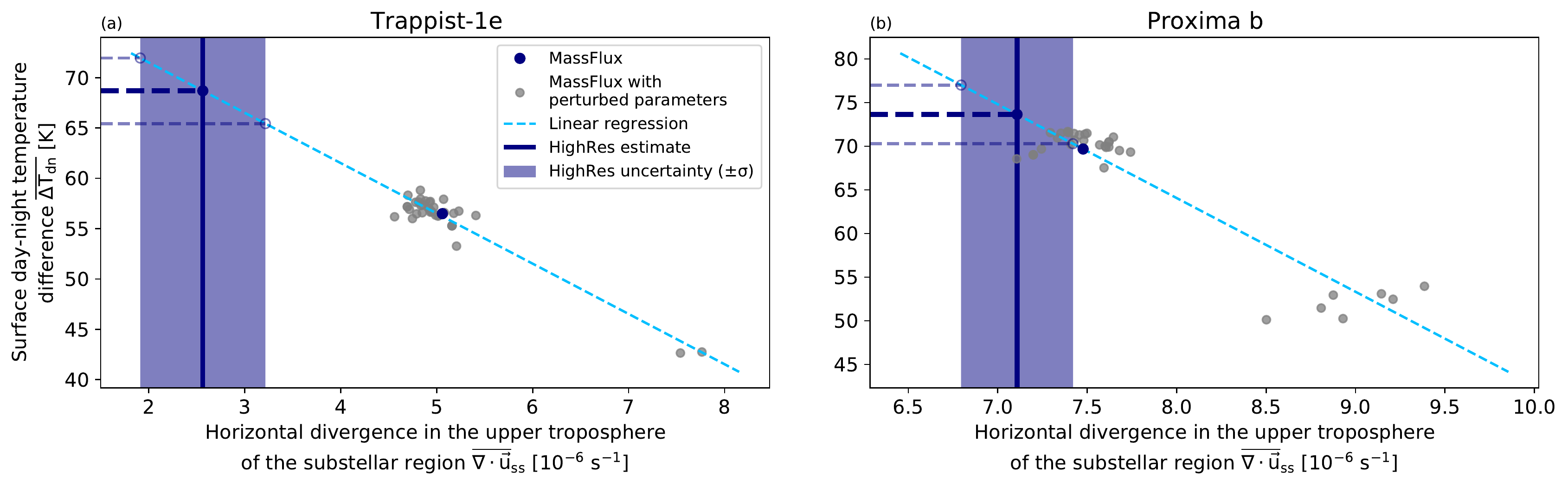}
\caption{Day-night surface temperature difference ($\overline{\Delta T_{dn}}$, \si{\kelvin}) as a function of the average wind divergence in the free troposphere (\SIrange{5}{20}{\km}) of the substellar region ($\overline{\nabla\cdot\vec{u}}_{ss}$, \si{\per\second}) for the ensemble of simulations with different empirical parameters of the mass-flux scheme (gray markers) in (a) Trappist-1e and (b) Proxima b set-up. The blue marker shows the value for the control simulation, i.e. \emph{MassFlux}. The dashed blue line shows linear fit, and the dark blue vertical lines show the divergence estimate from the \emph{HighRes} simulation, along with extrapolated $\overline{\Delta T_{dn}}$ (horizontal dashed lines). Dark blue shaded area represents the uncertainty in high-resolution estimates due to the temporal variability. \label{fig:regression}}
\end{figure*}

As has been shown in Sec.~\ref{sec:sens_conv}, the day-night temperature contrast is larger in the Proxima b than in the Trappist-1e case by \SI{\approx 15}{\kelvin} on average.
Perturbing the convection scheme parameters results in a scatter of the order of a several \si{\kelvin}, though with a few outliers with noticeably small $\overline{\Delta T_{dn}}$ (Fig.~\ref{fig:regression}).
With regards to the mean wind divergence over the substellar region, most simulations are clustered around \SI{5e-6}{\per\s} and \SI{7.5e-6}{\per\s} for Trappist-1e and Proxima b, respectively.
Increased divergence tends to correlate with smaller day-night temperature contrast, which is sensible since more vigorous outflow from the convective region is expected to redistribute heat around the planet more efficiently.
Extrapolating this relationship to the \emph{HighRes} simulation (vertical lines in Fig.~\ref{fig:regression}), we find that because of lower divergence, the estimate of $\overline{\Delta T_{dn}}$ is higher than in the global simulations.
The fact that \emph{HighRes} simulations produce weaker divergence than the global simulations is in agreement with the weaker latent heating in the upper troposphere, as has been illustrated by the vertical profiles in Fig.~\ref{fig:t_incr_vprof_lam}.
This is particularly clear for Trappist-1e, for which data from the \emph{HighRes} simulation predicts a temperature contrast of \SI{69\pm 3}{\kelvin}, i.e. an increase by almost \SI{12\pm 3}{\kelvin} compared to the control \emph{MassFlux} value (Fig.~\ref{fig:regression}a). 
As for Proxima b, the wind divergence in the \emph{HighRes} model is closer to the global model values.
Consequently, the increase of $\overline{\Delta T_{dn}}$ is only \SI{\approx 4}{\kelvin} relative to the global simulation.
This is to be expected from our earlier analysis, because cloud condensate profiles and TOA OLR histograms also showed a better match between the global and \emph{HighRes} models for Proxima b.
Still, it is evident from our high-resolution simulations that UM simulations with parameterized convection appear to be skewed towards higher upper-tropospheric divergence and thus smaller inter-hemispheric temperature contrasts.

\section{Discussion}
\subsection{Convection on the day side and implications for atmospheric circulation}
By controlling the magnitude of the substellar atmospheric heating on a tidally-locked terrestrial planet, convection clearly plays an important role in regulating its global climate.
Therefore, care should be taken when deciding between convection parameterizations to employ within a GCM.

On more massive and slower rotating planets, exemplified here by the Proxima b case, atmospheric circulation patterns can be expected to be less sensitive to how convection is represented.
Nevertheless, the thermodynamic aspects of climate would still be affected, namely a different global temperature and water vapor content, as well as a different cloud distribution.
For planets at the rotation regime boundary, exemplified by Trappist-1e, changing the convection scheme may tip the atmosphere into a markedly different circulation regime.
Simpler convection schemes, such as that based on adjustment or relaxation to a reference lapse rate used in this study, may lead to a positive bias in estimating mean global surface temperature, while underestimating the day-night temperature contrast (Sec.~\ref{sec:sens_conv}).
The latter is caused by exaggerated energy redistribution to the night side of the planet.
Whether this bias is positive for the majority of convection adjustment schemes is unclear and requires a broader model intercomparison project, such as \citet{FauchezEtAl2020}.

Focusing on the Trappist-1e case, which here is more sensitive to the convection representation, we note that convective heating, concentrated at the substellar point and represented differently by parameterizations, affects the global climate from three complementary perspectives.
First, it affects the thermal structure of the atmosphere, both on the day side and on the night side.
Second, it modulates the response of planetary-scale waves, favoring either equatorial or extratropical Rossby waves.
This is reflected by the configuration of dominant jet streams and the strength of equatorial super-rotation.
Third, it alters the global water balance, via changes in the total amount of water vapor, cloudiness, and the surface water redistribution.
The cloudiness response, expressed in terms of $CRE$, is substantial (see Table~\ref{tab:cre}), especially for stellar radiation.
The ``reflective'' effect of clouds is reduced by \SI{>60}{\percent} when a simpler convection scheme is used, causing the total (negative) $CRE$ to diminish.
This may have implications for the strength of the stabilizing cloud feedback and the estimates of the inner boundary of the habitable zone \citep{YangEtAl2013}.
Changes in the atmospheric circulation and radiation balance then feed back to the thermodynamic structure of the planet.
Thus, the effect of convection is multi-faceted.

The importance of developing convection parameterizations is not just to improve coarse-resolution GCMs.
More importantly, it helps us to disentangle the role of different processes in shaping planetary climates.
Parameterizations summarize our understanding of physical processes and their interactions with the large-scale flow.
Thus even with global cloud-permitting models becoming available, it is still valuable to revisit and improve convection parameterizations \citep{RioEtAl2019}.
Development of these parameterizations for exoplanets should be aided by convection-permitting models like the one in this study.

Convection-permitting models have been used to identify biases in global model parameterizations for Earth, for example in the case the structure, diurnal cycle, and rainfall intensity of the West African Monsoon \citep{MarshamEtAl2013}.
Whilst we think it is too early to compile similar guidelines for exoplanet GCMs --- due to a very wide parameter space and observational paucity --- in the present study we attempt to encourage further research in this direction by unveiling some of the benefits of high-resolution modeling.
Our high-resolution simulations illustrate this by producing a detailed representation of convective plumes in the substellar region, along with fine-scale cloud and precipitation patterns, supporting the results of \citet{ZhangEtAl2017} for a tidally-locked exoplanet.
Notwithstanding similar water vapor profiles and the same microphysics scheme used both in the global and \emph{HighRes} simulations, we find differences in the vertical profiles of latent heating and convective cloud condensate.
Coarse-resolution simulations with parameterized convection tend to have a positive bias in the latent heating profile, affecting the total temperature increment in the region.
We thus expect a weaker upper-troposphere heating if a hypothetical global convection-permitting simulation is performed.

Using horizontal flow divergence as one of the metrics of substellar convection, we make an estimate of what effect a hypothetical global convection-permitting simulation would have on the global climate, namely the mean day-night surface temperature contrast.
The results of our linear regression suggest that because of weaker divergence, the temperature contrast would be higher than in the coarse-resolution parameterized-convection experiments, especially in the Trappist-1e case.
This will be explored in more depth in our future work.

Finally, despite the thermal and circulation differences between convection regimes, our results show that the near-surface conditions remain relatively temperate.
Specifically, the day side of both planets always retains a patch of surface temperature above the freezing point of water.
At the same time, even in cold traps on the night side and even in the most extreme case (\emph{NoCnvPm} for Proxima b), the lowest temperature stays above the \ce{CO2} condensation point (\SI{\approx 125}{\kelvin} at $p_s=\SI{1}{\bar}$), thus keeping the contribution of this gas to the greenhouse effect.
If a high-resolution simulation is performed over the whole planet, we can expect the night side surface temperatures to be lower.
However, the decrease should not be large enough to go beyond the point of \ce{CO2} condensation.
Thus, given the atmospheric composition used in our study, it is still broadly possible for tidally-locked Trappist-1e and Proxima b to host a water cycle between hemispheres, confirming previous studies of these planets \citep{TurbetEtAl2016,BoutleEtAl2017,LewisEtAl2018,FauchezEtAl2020}.

\subsection{Implications for observations}
The cloudless atmosphere of a rocky planet can appear to have a large day-night temperature contrast, especially at low surface pressure \citep{Koll2019}.
The presence of condensible species that form clouds smooth this contrast via the release of latent heat of condensation.
As noted above, high clouds have a positive CRE by reducing the amount of energy radiated to space.
On tidally-locked planets, deep convective clouds, formed on the day side due to moist convection, reduce its thermal emission, thus further reducing the day-night contrast or even reversing it \citep{YangEtAl2013}.
Understanding the range of cloud cover uncertainty associated with different convection schemes can help us to interpret emission spectra that will be available from future space telescopes.

In our study, the effective temperature ($T_{eff}$) difference between the substellar and antistellar hemispheres varied from \SI{13}{\kelvin} to \SI{20}{\kelvin} for Trappist-1e and from \SI{24}{\kelvin} to \SI{35}{\kelvin} for Proxima b in global coarse-resolution experiments.
While the contrast is relatively small, it is still positive.
The high-resolution simulation tends to produce even lower thermal emission at the substellar point (Fig.~\ref{fig:toa_olr_hist}).
Thus running a global simulation with explicit convection may yield an even lower $T_{eff}$ contrast.

\subsection{Future work}
Whilst our study is one of the first attempts to use a convection-permitting model in the context of exoplanet atmospheres, its results raised several questions that are beyond the scope of the present manuscript.
To begin with, we used orbital parameters and stellar irradiation spectra for real exoplanets, but assumed synchronous rotation and a nitrogen-dominated atmosphere.
Our next step is to test the behavior of convection schemes at different surface pressures, atmospheric compositions (e.g. with oxygen and ozone), spin-orbit resonances, and stellar spectra of F-, G-, or K-stars.

From a technical perspective, it is also important to test the robustness of our conclusions by running a large suite of high-resolution simulations with various grid configurations and placements of the regional model domain.
This will provide more general guidelines for using convection schemes in exoplanet GCMs.
Ultimately, we aim to use the output of the convection-permitting model as a benchmark for refining GCM parameterizations of convection and microphysics.
Our future goal is to run a global convective-permitting simulation, which will be possible with the successor of the UM --- the LFRic model \citep{AdamsEtAl2019}.

The surface boundary condition is undoubtedly a very important part of modeling convection.
As shown by \citet{LewisEtAl2018}, placing a continent on the day side of Proxima b impedes surface evaporation, dries up the atmosphere, and reduces the heat redistribution to the night side.
These effects are clearly linked to convective processes, so future studies should also explore the response of high-resolution model to different types of surface, including various land types, a dynamic ocean, or an ocean with prescribed heat-flux.

Beyond thermodynamic effects of resolved convection for the climate of tidally-locked exoplanets, an exciting avenue of research would be to combine high-resolution modeling with interactive chemistry.
This would allow for a refined study of the gas mixing processes, such as the ozone cycle on Earth-like planets \citep{YatesEtAl2020} or the relative abundance of \ce{CH4} and \ce{CO} on hot Jupiters \citep{DrummondEtAl2020}.
More intense updrafts and downdrafts, as well as different rainfall rates, can lead to differences in how an exoplanet GCM simulates upward transport or deposition of chemical species.
Yet another unexplored problem, contingent on faithful simulations of atmospheric convection and clouds, as well as chemistry, is the occurrence of electric storms and generation of lightning \citep{TostEtAl2007}.
Because lightning can alter atmospheric composition on exoplanets \citep{ArdasevaEtAl2017}, high occurrence of electric storms of convective origin can manifest itself via a chemical footprint, such as nitrogen oxides or ozone.
Hence, high-resolution modeling can help to improve the interpretation of chemical signatures in observable spectra, by improving estimates of convection and electric storm activity.

\section{Conclusions}
\label{sec:conclusions}

The key findings of this study are as follows.
\begin{enumerate}
    \item Moist convection plays a key role in the heat redistribution on tidally-locked Earth-like exoplanets orbiting M-dwarf stars.
    By affecting the temperature and humidity stratification, convection modulates the global circulation and water cycle.
    Convection needs to be parameterized in global coarse-resolution models, but as our 3D GCM simulations demonstrate, one should keep in mind possible biases of simplified parameterizations.
    Swapping a mass-flux convection scheme to a simple adjustment scheme can lead to a substantial decrease of the shortwave cloud radiative effect (by \SI{>60}{\percent}) and albedo of the planet (by \SI{>50}{\percent}), potentially shrinking the width of habitable zone for a given star.
    \item Importantly, convection regulates the climate of the whole planet and not just the substellar point where it is active.
    This is linked to the change in circulation patterns for Trappist-1e and thermodynamic effects for Proxima b, when one or another convection parameterization is chosen.
    The convective adjustment parameterization warms up the night-side cold traps by \SIrange{17}{36}{\K} for these planets relative to the mass-flux scheme.
    The overall top-of-atmosphere thermal contrast between the substellar and antistellar hemispheres is altered too: from \SI{13}{\kelvin} to \SI{20}{\kelvin} for Trappist-1e and from \SI{24}{\kelvin} to \SI{35}{\kelvin} for Proxima b.
    \item Convection parameterization can have different effects for different terrestrial planets.
    Even with the minor orbital differences used here for Trappist-1e and Proxima b, we find that the former is more susceptible to changing the parameterization.
    We speculate that this is because the atmosphere of Trappist-1e is on the boundary between two rotation regimes, and so can be more easily tipped over into one dynamical state or another.
    \item From the habitability perspective, the near-surface conditions both in the Trappist-1e and Proxima b cases remain temperate and allow for a water cycle to exist between the hemispheres.
    In other words, the choice of convective treatments we have explored do not push the climate out of a habitable state, given the atmospheric composition and planetary parameters used in our study.
    \item GCMs with convection parameterizations may suffer from certain biases compared to explicitly resolved convection models.
    In our case, the global UM is biased in the cloud structure and cloud radiative effects, relative to the high-resolution model.
    \item We perform an ensemble of global simulations to predict what effect a hypothetical global convection-permitting simulation would have on the global climate, namely the mean day-night surface temperature contrast.
    Our results suggest that in such a simulation the temperature contrast would be higher than in the coarse-resolution simulations with parameterized convection, by \SI{\approx 12}{\K} for Trappist-1e and by \SI{\approx 4}{\K} for Proxima b.
    \item A novel approach applied here is convection-permitting modeling of substellar convection.
    We use the nested grid configuration of the UM, establishing a self-consistent set-up between the global coarse-resolution and regional high-resolution configurations.
    We show that convection-permitting simulations can be used to assess convection parameterizations in GCMs and their impact on the climate of an exoplanet as a whole.
    Future convection-permitting simulations will help to constrain the amount of cloud, precipitation, and export of moisture to the night side of wider range of tidally-locked exoplanets.
\end{enumerate}

\acknowledgments
This work was partly supported by a Science and Technology Facilities Council Consolidated Grant (\texttt{ST/R000395/1}).
We acknowledge use of the MONSooN system, a collaborative facility supplied under the Joint Weather and Climate Research Programme, a strategic partnership between the Met Office and the Natural Environment Research Council.
This work was performed using the DiRAC Data Intensive service at Leicester, operated by the University of Leicester IT Services, which forms part of the STFC DiRAC HPC Facility (\url{www.dirac.ac.uk}). The equipment was funded by BEIS capital funding via STFC capital grants \texttt{ST/K000373/1} and \texttt{ST/R002363/1} and STFC DiRAC Operations grant \texttt{ST/R001014/1}. DiRAC is part of the National e-Infrastructure.

\software{
Python code used in this study is publicly available, including the library for model output post-processing (\url{https://github.com/exoclim/aeolus}), scripts to generate the figures (\url{https://github.com/dennissergeev/exoconvection-apj-2020}), and their dependencies:
iris \citep{iris},
matplotlib \citep{matplotlib},
pyvista \citep{pyvista}.
}

\appendix
\section{Convection scheme parameters in the global ensemble simulations}
In order to derive the statistical relationship between the substellar convection and global temperature conditions, we run a suite of global coarse-resolution experiments with some of the parameters of the mass-flux convection scheme perturbed within a range of values (see Sec.~\ref{sec:dayside_impact}).
Only a single parameter is perturbed in each realization of the ensemble, the rest are held the same as in the control.
Here we list these parameters, their value in the control experiment, and range of values in the ensemble experiments.

While this is not a complete inventory of all parameters available in the UM, it includes the most important ones.
The first five parameters regulate the rate of entrainment of background air into the convective plume and detrainment of convective clouds, which have been shown to be key factors in reproducing tropical atmosphere soundings \citep{DerbyshireEtAl2004} and simulating convective aggregation in aquaplanet experiments \citep{BeckerEtAl2017}.
For the mass-flux parameterization used in the UM specifically, \citet{SextonEtAl2019} show that increased values of \texttt{ent\_fac\_dp} lead to the reduction of convection depth and suppression of precipitation.
Larger values of \texttt{r\_det}, on the other hand, give deeper convection and also change its vertical profile.
For the CAPE timescale, for example, \citet{SextonEtAl2019} report that shorter values result in more spatially and temporally intermittent convection.
We also briefly looked at the simulation with convective momentum transport switched off (\texttt{l\_mom}$=OFF$), and it appeared to be one of the outliers in Fig.~\ref{fig:regression}, which warrants further sensitivity studies.
Other parameters appear to be important too, though to a lesser extent.

\begin{deluxetable*}{llll}
\tablecaption{Summary of parameters in the mass-flux convection scheme and their values in ensemble experiments.\label{tab:grcs_ens}}
\tablewidth{0pt}
\tablehead{
\colhead{Parameter} & \colhead{Control value} & \colhead{Perturbed values} & \colhead{Comments}}
\startdata
\texttt{ent\_opt\_dp}          & $3$                  & $0$, $1$, $2$, $4$, $5$, $6$, $7$                       & Entrainment amplitude profile for deep convection \\
\texttt{ent\_fac\_dp}          & $1.13$               & $10^{-2}$, $0.1$, $0.5$, $2.0$                          & Entrainment factor for deep convection \\
\texttt{ent\_fac\_md}          & $0.9$                & $10^{-2}$, $0.1$, $0.5$, $2.0$                          & Entrainment factor for mid-level convection \\
\texttt{amdet\_fac}            & $3.0$                & $10^{-2}$, $0.1$, $10.0$, $20.0$                        & Mixing detrainment parameter \\
\texttt{r\_det}                & $0.8$                & $0.1$, $0.3$, $0.6$, $1$                                & Adaptive detrainment coefficient \\
\texttt{cape\_timescale}       & $3600$               & $300$, $1800$, $14400$, $28800$                         & $e$-folding time for the dissipation of CAPE (\si{\s}) \\
\texttt{w\_cape\_limit}        & $0.4$                & $10^{-2}$, $0.1$, $1$, $10.0$                           & CAPE threshold for vertical velocity \\
\texttt{cnv\_wat\_load\_opt}   & OFF                  & ON                                                      & Effects of water loading in calculation of buoyancy \\
\texttt{mid\_cnv\_pmin}        & $100$                & $10$                                                    & Minimum pressure for mid-level convection (\si{\hecto\pascal}) \\
\texttt{mparwtr}               & $10^{-3}$            & $10^{-5}$, $10^{-4}$, $10^{-2}$                         & Maximum value of the critical cloud condensate profile \\
\texttt{qlmin}                 & $2\times10^{-4}$     & $10^{-8}$, $10^{-6}$, $10^{-3}$                         & Minimum value of the critical cloud condensate profile \\
\texttt{fac\_qsat}             & $0.5$                & $10^{-2}$, $0.1$, $5.0$                                 & Saturation water vapor factor for cloud condensate \\
\texttt{l\_mom}                & ON                   & OFF                                                     & Convective momentum transport \\
\texttt{l\_mom\_dd}            & OFF                  & ON                                                      & Convective momentum transport in downdrafts \\
\texttt{cnv\_cold\_pools}      & OFF                  & Propagating, Single-column                              & Cloud pools generated by downdrafts \\
\enddata
\tablecomments{For detailed description of these parameters see the UM documentation.}
\end{deluxetable*}

\bibliography{references}{}

\begin{thebibliography}{}
\expandafter\ifx\csname natexlab\endcsname\relax\def\natexlab#1{#1}\fi
\providecommand{\url}[1]{\href{#1}{#1}}
\providecommand{\dodoi}[1]{doi:~\href{http://doi.org/#1}{\nolinkurl{#1}}}
\providecommand{\doeprint}[1]{\href{http://ascl.net/#1}{\nolinkurl{http://ascl.net/#1}}}
\providecommand{\doarXiv}[1]{\href{https://arxiv.org/abs/#1}{\nolinkurl{https://arxiv.org/abs/#1}}}

\bibitem[{Adams {et~al.}(2019)Adams, Ford, Hambley, Hobson, Kav{\v{c}}i{\v{c}},
  Maynard, Melvin, M{\"{u}}ller, Mullerworth, Porter, Rezny, Shipway, \&
  Wong}]{AdamsEtAl2019}
Adams, S.~V., Ford, R.~W., Hambley, M., {et~al.} 2019, Journal of Parallel and
  Distributed Computing, 132, 383, \dodoi{10.1016/j.jpdc.2019.02.007}

\bibitem[{Anglada-Escud{\'{e}} {et~al.}(2016)Anglada-Escud{\'{e}}, Amado,
  Barnes, Berdi{\~{n}}as, Butler, Coleman, De~La~Cueva, Dreizler, Endl,
  Giesers, Jeffers, Jenkins, Jones, Kiraga, K{\"{u}}rster,
  L{\'{o}}pez-Gonz{\'{a}}lez, Marvin, Morales, Morin, Nelson, Ortiz, Ofir,
  Paardekooper, Reiners, Rodr{\'{i}}guez, Rodr{\'{i}}guez-L{\'{o}}pez,
  Sarmiento, Strachan, Tsapras, Tuomi, \& Zechmeister}]{Anglada-Escude2016}
Anglada-Escud{\'{e}}, G., Amado, P.~J., Barnes, J., {et~al.} 2016, Nature, 536,
  437, \dodoi{10.1038/nature19106}

\bibitem[{Arakawa(2004)}]{Arakawa2004}
Arakawa, A. 2004, Journal of Climate, 17, 2493,
  \dodoi{10.1175/1520-0442(2004)017<2493:RATCPP>2.0.CO;2}

\bibitem[{Ardaseva {et~al.}(2017)Ardaseva, Rimmer, Waldmann, Rocchetto,
  Yurchenko, Helling, \& Tennyson}]{ArdasevaEtAl2017}
Ardaseva, A., Rimmer, P.~B., Waldmann, I., {et~al.} 2017, Monthly Notices of
  the Royal Astronomical Society, 470, 187, \dodoi{10.1093/mnras/stx1012}

\bibitem[{Becker {et~al.}(2017)Becker, Stevens, \& Hohenegger}]{BeckerEtAl2017}
Becker, T., Stevens, B., \& Hohenegger, C. 2017, Journal of Advances in
  Modeling Earth Systems, 9, 1488, \dodoi{10.1002/2016MS000865}

\bibitem[{Betts(1986)}]{Betts1986}
Betts, A.~K. 1986, Quarterly Journal of the Royal Meteorological Society, 112,
  677, \dodoi{10.1002/qj.49711247307}

\bibitem[{Boutle {et~al.}(2017)Boutle, Mayne, Drummond, Manners, Goyal,
  Hugo~Lambert, Acreman, \& Earnshaw}]{BoutleEtAl2017}
Boutle, I.~A., Mayne, N.~J., Drummond, B., {et~al.} 2017, Astronomy {\&}
  Astrophysics, 601, A120, \dodoi{10.1051/0004-6361/201630020}

\bibitem[{Bretherton \& Khairoutdinov(2015)}]{BrethertonKhairoutdinov2015}
Bretherton, C.~S., \& Khairoutdinov, M.~F. 2015, Journal of Advances in
  Modeling Earth Systems, 7, 1765, \dodoi{10.1002/2015MS000499}

\bibitem[{Bush {et~al.}(2019)Bush, Allen, Bain, Boutle, Edwards, Finnenkoetter,
  Franklin, Hanley, Lean, Lock, Manners, Mittermaier, Morcrette, North, Petch,
  Short, Vosper, Walters, Webster, Weeks, Wilkinson, Wood, \&
  Zerroukat}]{BushEtAl2019}
Bush, M., Allen, T., Bain, C., {et~al.} 2019, Geoscientific Model Development
  Discussions, 1, \dodoi{10.5194/gmd-2019-130}

\bibitem[{Carone {et~al.}(2015)Carone, Keppens, \& Decin}]{CaroneEtAl2015}
Carone, L., Keppens, R., \& Decin, L. 2015, Monthly Notices of the Royal
  Astronomical Society, 453, 2413, \dodoi{10.1093/mnras/stv1752}

\bibitem[{Ceppi {et~al.}(2017)Ceppi, Brient, Zelinka, \&
  Hartmann}]{CeppiEtAl2017}
Ceppi, P., Brient, F., Zelinka, M.~D., \& Hartmann, D.~L. 2017, Wiley
  Interdisciplinary Reviews: Climate Change, 8, e465, \dodoi{10.1002/wcc.465}

\bibitem[{{Del Genio} {et~al.}(2018){Del Genio}, Way, Amundsen, Aleinov,
  Kelley, Kiang, \& Clune}]{DelGenioEtAl2018}
{Del Genio}, A.~D., Way, M.~J., Amundsen, D.~S., {et~al.} 2018, Astrobiology,
  ast.2017.1760, \dodoi{10.1089/ast.2017.1760}

\bibitem[{Derbyshire {et~al.}(2004)Derbyshire, Beau, Bechtold, Grandpeix,
  Piriou, Redelsperger, \& Soares}]{DerbyshireEtAl2004}
Derbyshire, S., Beau, I., Bechtold, P., {et~al.} 2004, Quarterly Journal of the
  Royal Meteorological Society, 130, 3055, \dodoi{10.1256/qj.03.130}

\bibitem[{Dressing \& Charbonneau(2015)}]{DressingCharbonneau2015}
Dressing, C.~D., \& Charbonneau, D. 2015, The Astrophysical Journal, 807, 45,
  \dodoi{10.1088/0004-637X/807/1/45}

\bibitem[{Drummond {et~al.}(2020)Drummond, Hebrard, Mayne, Venot, Ridgway,
  Changeat, Tsai, Manners, Tremblin, Abraham, Sing, \&
  Kohary}]{DrummondEtAl2020}
Drummond, B., Hebrard, E., Mayne, N.~J., {et~al.} 2020, Astronomy {\&}
  Astrophysics.
\newblock \url{http://arxiv.org/abs/2001.11444}

\bibitem[{Fauchez {et~al.}(2020)Fauchez, Turbet, Wolf, Boutle, Way, Del~Genio,
  Mayne, Tsigaridis, Kopparapu, Yang, Forget, Mandell, \&
  Domagal~Goldman}]{FauchezEtAl2020}
Fauchez, T.~J., Turbet, M., Wolf, E.~T., {et~al.} 2020, Geosci. Model Dev, 13,
  707, \dodoi{10.5194/gmd-13-707-2020}

\bibitem[{Frierson(2007)}]{Frierson2007}
Frierson, D. M.~W. 2007, Journal of the Atmospheric Sciences, 64, 1959,
  \dodoi{10.1175/JAS3935.1}

\bibitem[{Frierson {et~al.}(2006)Frierson, Held, \&
  Zurita-Gotor}]{Frierson2006}
Frierson, D. M.~W., Held, I.~M., \& Zurita-Gotor, P. 2006, Journal of the
  Atmospheric Sciences, 63, 2548, \dodoi{10.1175/JAS3753.1}

\bibitem[{Gillon {et~al.}(2017)Gillon, Triaud, Demory, Jehin, Agol, Deck,
  Lederer, de~Wit, Burdanov, Ingalls, Bolmont, Leconte, Raymond, Selsis,
  Turbet, Barkaoui, Burgasser, Burleigh, Carey, Chaushev, Copperwheat, Delrez,
  Fernandes, Holdsworth, Kotze, Van~Grootel, Almleaky, Benkhaldoun, Magain, \&
  Queloz}]{GillonEtAl2017}
Gillon, M., Triaud, A. H. M.~J., Demory, B.-O., {et~al.} 2017, Nature, 542,
  456, \dodoi{10.1038/nature21360}

\bibitem[{Goldblatt(2015)}]{Goldblatt2015}
Goldblatt, C. 2015, Astrobiology, 15, 362, \dodoi{10.1089/ast.2014.1268}

\bibitem[{G{\'{o}}mez-Leal {et~al.}(2019)G{\'{o}}mez-Leal, Kaltenegger,
  Lucarini, \& Lunkeit}]{Gomez-Leal2019}
G{\'{o}}mez-Leal, I., Kaltenegger, L., Lucarini, V., \& Lunkeit, F. 2019,
  Icarus, 321, 608, \dodoi{10.1016/J.ICARUS.2018.11.019}

\bibitem[{Gregory \& Rowntree(1990)}]{GregoryRowntree1990}
Gregory, D., \& Rowntree, P.~R. 1990, {A Mass Flux Convection Scheme with
  Representation of Cloud Ensemble Characteristics and Stability-Dependent
  Closure}, \dodoi{10.1175/1520-0493(1990)118<1483:AMFCSW>2.0.CO;2}

\bibitem[{Grimm {et~al.}(2018)Grimm, Demory, Gillon, Dorn, Agol, Burdanov,
  Delrez, Sestovic, Triaud, Turbet, Bolmont, Caldas, de~Wit, Jehin, Leconte,
  Raymond, Van~Grootel, Burgasser, Carey, Fabrycky, Heng, Hernandez, Ingalls,
  Lederer, Selsis, \& Queloz}]{Grimm2018}
Grimm, S.~L., Demory, B.-O., Gillon, M., {et~al.} 2018, Astronomy {\&}
  Astrophysics, 613, A68, \dodoi{10.1051/0004-6361/201732233}

\bibitem[{Hammond \& Pierrehumbert(2018)}]{HammondPierrehumbert2018}
Hammond, M., \& Pierrehumbert, R.~T. 2018, The Astrophysical Journal, 869, 65,
  \dodoi{10.3847/1538-4357/aaec03}

\bibitem[{Haqq-Misra {et~al.}(2017)Haqq-Misra, Wolf, Joshi, Zhang, \&
  Kopparapu}]{Haqq-MisraEtAl2017}
Haqq-Misra, J., Wolf, E.~T., Joshi, M., Zhang, X., \& Kopparapu, R.~K. 2017,
  The Astrophysical Journal, 852, 67, \dodoi{10.3847/1538-4357/aa9f1f}

\bibitem[{Hunter(2007)}]{matplotlib}
Hunter, J.~D. 2007, Computing in Science \& Engineering, 9, 90,
  \dodoi{10.1109/MCSE.2007.55}

\bibitem[{Joshi {et~al.}(2020)Joshi, Elvidge, Wordsworth, \&
  Sergeev}]{JoshiEtAl2020}
Joshi, M.~M., Elvidge, A.~D., Wordsworth, R., \& Sergeev, D. 2020, The
  Astrophysical Journal Letters, 892, L33, \dodoi{10.3847/2041-8213/ab7fb3}

\bibitem[{Kasting {et~al.}(2014)Kasting, Kopparapu, Ramirez, \&
  Harman}]{KastingEtAl2014}
Kasting, J.~F., Kopparapu, R., Ramirez, R.~M., \& Harman, C.~E. 2014,
  Proceedings of the National Academy of Sciences, 111, 12641,
  \dodoi{10.1073/pnas.1309107110}

\bibitem[{Kasting {et~al.}(1993)Kasting, Whitmire, \&
  Reynolds}]{KastingEtAl1993}
Kasting, J.~F., Whitmire, D.~P., \& Reynolds, R.~T. 1993, Icarus, 101, 108,
  \dodoi{10.1006/icar.1993.1010}

\bibitem[{Kiladis {et~al.}(2009)Kiladis, Wheeler, Haertel, Straub, \&
  Roundy}]{KiladisEtAl2009}
Kiladis, G.~N., Wheeler, M.~C., Haertel, P.~T., Straub, K.~H., \& Roundy, P.~E.
  2009, Reviews of Geophysics, 47, \dodoi{10.1029/2008RG000266}

\bibitem[{Koll(2019)}]{Koll2019}
Koll, D. D.~B. 2019, Astrophysical Journal.
\newblock \url{http://arxiv.org/abs/1907.13145}

\bibitem[{Komacek \& Abbot(2019)}]{KomacekAbbot2019}
Komacek, T.~D., \& Abbot, D.~S. 2019, The Astrophysical Journal, 871, 245,
  \dodoi{10.3847/1538-4357/aafb33}

\bibitem[{Kopparapu {et~al.}(2017)Kopparapu, Wolf, Arney, Batalha, Haqq-Misra,
  Grimm, \& Heng}]{KopparapuEtAl2017}
Kopparapu, R.~K., Wolf, E.~T., Arney, G., {et~al.} 2017, The Astrophysical
  Journal, 845, 5, \dodoi{10.3847/1538-4357/aa7cf9}

\bibitem[{Lambert {et~al.}(2020)Lambert, Challenor, Lewis, McNeall, Owen,
  Boutle, Christensen, Keane, Stirling, Webb, \& Mayne}]{LambertEtAl2020}
Lambert, F.~H., Challenor, P., Lewis, N.~T., {et~al.} 2020,
  \dodoi{10.1002/ESSOAR.10502338.1}

\bibitem[{Leconte {et~al.}(2013)Leconte, Forget, Charnay, Wordsworth, Selsis,
  Millour, \& Spiga}]{LeconteEtAl2013}
Leconte, J., Forget, F., Charnay, B., {et~al.} 2013, Astronomy {\&}
  Astrophysics, 554, A69, \dodoi{10.1051/0004-6361/201321042}

\bibitem[{Leconte {et~al.}(2015)Leconte, Wu, Menou, \&
  Murray}]{LeconteEtAl2015}
Leconte, J., Wu, H., Menou, K., \& Murray, N. 2015, Science, 347, 632,
  \dodoi{10.1126/science.1258686}

\bibitem[{Lef{\`{e}}vre {et~al.}(2018)Lef{\`{e}}vre, Lebonnois, \&
  Spiga}]{LefevreEtAl2018}
Lef{\`{e}}vre, M., Lebonnois, S., \& Spiga, A. 2018, Journal of Geophysical
  Research: Planets, 123, 2773, \dodoi{10.1029/2018JE005679}

\bibitem[{Lewis {et~al.}(2018)Lewis, Lambert, Boutle, Mayne, Manners, \&
  Acreman}]{LewisEtAl2018}
Lewis, N.~T., Lambert, F.~H., Boutle, I.~A., {et~al.} 2018, The Astrophysical
  Journal, 854, 171, \dodoi{10.3847/1538-4357/aaad0a}

\bibitem[{Lines {et~al.}(2018)Lines, Manners, Mayne, Goyal, Carter, Boutle,
  Lee, Helling, Drummond, Acreman, \& Sing}]{LinesEtAl2018b}
Lines, S., Manners, J., Mayne, N.~J., {et~al.} 2018, Monthly Notices of the
  Royal Astronomical Society, 481, 194, \dodoi{10.1093/mnras/sty2275}

\bibitem[{Maher {et~al.}(2018)Maher, Vallis, Sherwood, Webb, \&
  Sansom}]{MaherEtAl2018}
Maher, P., Vallis, G.~K., Sherwood, S.~C., Webb, M.~J., \& Sansom, P.~G. 2018,
  Geophysical Research Letters, 45, 3728, \dodoi{10.1002/2017GL076826}

\bibitem[{Manabe {et~al.}(1965)Manabe, Smagorinsky, \&
  Strickler}]{ManabeEtAl1965}
Manabe, S., Smagorinsky, J., \& Strickler, R.~F. 1965, Monthly Weather Review,
  93, 769, \dodoi{10.1175/1520-0493(1965)093<0769:scoagc>2.3.co;2}

\bibitem[{Marsham {et~al.}(2013)Marsham, Dixon, Garcia-Carreras, Lister,
  Parker, Knippertz, \& Birch}]{MarshamEtAl2013}
Marsham, J.~H., Dixon, N.~S., Garcia-Carreras, L., {et~al.} 2013, Geophysical
  Research Letters, 40, 1843, \dodoi{10.1002/grl.50347}

\bibitem[{Mayne {et~al.}(2014{\natexlab{a}})Mayne, Baraffe, Acreman, Smith,
  Wood, Amundsen, Thuburn, \& Jackson}]{MayneEtAl2014b}
Mayne, N.~J., Baraffe, I., Acreman, D.~M., {et~al.} 2014{\natexlab{a}}, Geosci.
  Model Dev, 7, 3059, \dodoi{10.5194/gmd-7-3059-2014}

\bibitem[{Mayne {et~al.}(2019)Mayne, Drummond, Debras, Jaupart, Manners,
  Boutle, Baraffe, \& Kohary}]{MayneEtAl2019}
Mayne, N.~J., Drummond, B., Debras, F., {et~al.} 2019, The Astrophysical
  Journal, 871, 56, \dodoi{10.3847/1538-4357/aaf6e9}

\bibitem[{Mayne {et~al.}(2014{\natexlab{b}})Mayne, Baraffe, Acreman, Smith,
  Browning, Amundsen, Wood, Thuburn, \& Jackson}]{MayneEtAl2014a}
Mayne, N.~J., Baraffe, I., Acreman, D.~M., {et~al.} 2014{\natexlab{b}},
  Astronomy {\&} Astrophysics, 561, A1, \dodoi{10.1051/0004-6361/201322174}

\bibitem[{Merlis \& Schneider(2010)}]{MerlisSchneider2010}
Merlis, T.~M., \& Schneider, T. 2010, Journal of Advances in Modeling Earth
  Systems, 2, 1, \dodoi{10.3894/JAMES.2010.2.13}

\bibitem[{{Met Office}(2010-2020)}]{iris}
{Met Office}. 2010-2020, {Iris: A Python library for analysing and visualising
  meteorological and oceanographic data sets}.
\newblock \url{http://scitools.org.uk/}

\bibitem[{Mitchell {et~al.}(2009)Mitchell, Pierrehumbert, Frierson, \&
  Caballero}]{MitchellEtAl2009}
Mitchell, J.~L., Pierrehumbert, R.~T., Frierson, D.~M., \& Caballero, R. 2009,
  Icarus, 203, 250, \dodoi{10.1016/J.ICARUS.2009.03.043}

\bibitem[{Mitchell \& Vallis(2010)}]{MitchellVallis2010}
Mitchell, J.~L., \& Vallis, G.~K. 2010, Journal of Geophysical Research, 115,
  E12008, \dodoi{10.1029/2010JE003587}

\bibitem[{Penn \& Vallis(2018)}]{PennVallis2018}
Penn, J., \& Vallis, G.~K. 2018, The Astrophysical Journal, 868, 147,
  \dodoi{10.3847/1538-4357/aaeb20}

\bibitem[{Rajpurohit {et~al.}(2013)Rajpurohit, Reyl{\'{e}}, Allard, Homeier,
  Schultheis, Bessell, \& Robin}]{RajpurohitEtAl2013}
Rajpurohit, A.~S., Reyl{\'{e}}, C., Allard, F., {et~al.} 2013, Astronomy and
  Astrophysics, 556, \dodoi{10.1051/0004-6361/201321346}

\bibitem[{Rio {et~al.}(2019)Rio, Del~Genio, \& Hourdin}]{RioEtAl2019}
Rio, C., Del~Genio, A.~D., \& Hourdin, F. 2019, Current Climate Change Reports,
  5, 95, \dodoi{10.1007/s40641-019-00127-w}

\bibitem[{Sergeev {et~al.}(2017)Sergeev, Renfrew, Spengler, \&
  Dorling}]{SergeevEtAl2017}
Sergeev, D.~E., Renfrew, I.~A., Spengler, T., \& Dorling, S.~R. 2017, Quarterly
  Journal of the Royal Meteorological Society, 143, 12, \dodoi{10.1002/qj.2911}

\bibitem[{Sexton {et~al.}(2019)Sexton, Karmalkar, Murphy, Williams, Boutle,
  Morcrette, Stirling, \& Vosper}]{SextonEtAl2019}
Sexton, D. M.~H., Karmalkar, A.~V., Murphy, J.~M., {et~al.} 2019, Climate
  Dynamics, 1, \dodoi{10.1007/s00382-019-04625-3}

\bibitem[{Sherwood {et~al.}(2014)Sherwood, Bony, \&
  Dufresne}]{SherwoodEtAl2014}
Sherwood, S.~C., Bony, S., \& Dufresne, J.-L. 2014, Nature, 505, 37,
  \dodoi{10.1038/nature12829}

\bibitem[{Showman \& Guillot(2002)}]{ShowmanGuillot2002}
Showman, A.~P., \& Guillot, T. 2002, Astronomy {\&} Astrophysics, 385, 166,
  \dodoi{10.1051/0004-6361:20020101}

\bibitem[{Showman \& Polvani(2011)}]{ShowmanPolvani2011}
Showman, A.~P., \& Polvani, L.~M. 2011, The Astrophysical Journal, 738, 71,
  \dodoi{10.1088/0004-637X/738/1/71}

\bibitem[{Spiga {et~al.}(2017)Spiga, Hinson, Madeleine, Navarro, Millour,
  Forget, \& Montmessin}]{SpigaEtAl2017}
Spiga, A., Hinson, D.~P., Madeleine, J.~B., {et~al.} 2017, Nature Geoscience,
  10, 652, \dodoi{10.1038/ngeo3008}

\bibitem[{Stratton {et~al.}(2018)Stratton, Senior, Vosper, Folwell, Boutle,
  Earnshaw, Kendon, Lock, Malcolm, Manners, Morcrette, Short, Stirling, Taylor,
  Tucker, Webster, \& Wilkinson}]{StrattonEtAl2018}
Stratton, R.~A., Senior, C.~A., Vosper, S.~B., {et~al.} 2018, Journal of
  Climate, 31, 3485, \dodoi{10.1175/JCLI-D-17-0503.1}

\bibitem[{Sullivan \& Kaszynski(2019)}]{pyvista}
Sullivan, C., \& Kaszynski, A. 2019, Journal of Open Source Software, 4, 1450,
  \dodoi{10.21105/joss.01450}

\bibitem[{Thomson \& Vallis(2019)}]{ThomsonVallis2019}
Thomson, S.~I., \& Vallis, G.~K. 2019, Atmosphere, 10, 803,
  \dodoi{10.3390/atmos10120803}

\bibitem[{Tost {et~al.}(2007)Tost, J{\"{o}}ckel, \& Lelieveld}]{TostEtAl2007}
Tost, H., J{\"{o}}ckel, P., \& Lelieveld, J. 2007, Atmospheric Chemistry and
  Physics, 7, 4553, \dodoi{10.5194/acp-7-4553-2007}

\bibitem[{Turbet {et~al.}(2016)Turbet, Leconte, Selsis, Bolmont, Forget, Ribas,
  Raymond, \& Anglada-Escud{\'{e}}}]{TurbetEtAl2016}
Turbet, M., Leconte, J., Selsis, F., {et~al.} 2016, Astronomy {\&}
  Astrophysics, 596, A112, \dodoi{10.1051/0004-6361/201629577}

\bibitem[{Turbet {et~al.}(2018)Turbet, Bolmont, Leconte, Forget, Selsis, Tobie,
  Caldas, Naar, \& Gillon}]{TurbetEtAl2018}
Turbet, M., Bolmont, E., Leconte, J., {et~al.} 2018, Astronomy {\&}
  Astrophysics, 612, A86, \dodoi{10.1051/0004-6361/201731620}

\bibitem[{Walters {et~al.}(2019)Walters, Baran, Boutle, Brooks, Earnshaw,
  Edwards, Furtado, Hill, Lock, Manners, Morcrette, Mulcahy, Sanchez, Smith,
  Stratton, Tennant, Tomassini, Van~Weverberg, Vosper, Willett, Browse,
  Bushell, Carslaw, Dalvi, Essery, Gedney, Hardiman, Johnson, Johnson, Jones,
  Jones, Mann, Milton, Rumbold, Sellar, Ujiie, Whitall, Williams, \&
  Zerroukat}]{WaltersEtAl2019}
Walters, D., Baran, A.~J., Boutle, I., {et~al.} 2019, Geoscientific Model
  Development, 12, 1909, \dodoi{10.5194/gmd-12-1909-2019}

\bibitem[{Way {et~al.}(2018)Way, Del~Genio, Aleinov, Clune, Kelley, \&
  Kiang}]{WayEtAl2018a}
Way, M.~J., Del~Genio, A.~D., Aleinov, I., {et~al.} 2018, The Astrophysical
  Journal Supplement Series, 239, 24, \dodoi{10.3847/1538-4365/aae9e1}

\bibitem[{Way {et~al.}(2017)Way, Aleinov, Amundsen, Chandler, Clune, Genio,
  Fujii, Kelley, Kiang, Sohl, \& Tsigaridis}]{WayEtAl2017}
Way, M.~J., Aleinov, I., Amundsen, D.~S., {et~al.} 2017, The Astrophysical
  Journal Supplement Series, 231, 12, \dodoi{10.3847/1538-4365/aa7a06}

\bibitem[{Wing {et~al.}(2018)Wing, Reed, Satoh, Stevens, Bony, \&
  Ohno}]{WingEtAl2018}
Wing, A.~A., Reed, K.~A., Satoh, M., {et~al.} 2018, Geosci. Model Dev, 11, 793,
  \dodoi{10.5194/gmd-11-793-2018}

\bibitem[{Wolf(2017)}]{Wolf2017}
Wolf, E.~T. 2017, The Astrophysical Journal, 839, L1,
  \dodoi{10.3847/2041-8213/aa693a}

\bibitem[{Wolf \& Toon(2015)}]{WolfToon2015}
Wolf, E.~T., \& Toon, O.~B. 2015, Journal of Geophysical Research: Atmospheres,
  120, 5775, \dodoi{10.1002/2015JD023302}

\bibitem[{Wood {et~al.}(2014)Wood, Staniforth, White, Allen, Diamantakis,
  Gross, Melvin, Smith, Vosper, Zerroukat, \& Thuburn}]{WoodEtAl2014}
Wood, N., Staniforth, A., White, A., {et~al.} 2014, Quarterly Journal of the
  Royal Meteorological Society, 140, 1505, \dodoi{10.1002/qj.2235}

\bibitem[{Wordsworth(2015)}]{Wordsworth2015}
Wordsworth, R. 2015, The Astrophysical Journal, 806, 180,
  \dodoi{10.1088/0004-637X/806/2/180}

\bibitem[{Wordsworth {et~al.}(2011)Wordsworth, Forget, Selsis, Millour,
  Charnay, \& Madeleine}]{WordsworthEtAl2011}
Wordsworth, R.~D., Forget, F., Selsis, F., {et~al.} 2011, Astrophysical Journal
  Letters, 733, \dodoi{10.1088/2041-8205/733/2/L48}

\bibitem[{Yang \& Abbot(2014)}]{YangAbbot2014}
Yang, J., \& Abbot, D.~S. 2014, The Astrophysical Journal, 784, 155,
  \dodoi{10.1088/0004-637X/784/2/155}

\bibitem[{Yang {et~al.}(2013)Yang, Cowan, \& Abbot}]{YangEtAl2013}
Yang, J., Cowan, N.~B., \& Abbot, D.~S. 2013, The Astrophysical Journal, 771,
  L45, \dodoi{10.1088/2041-8205/771/2/L45}

\bibitem[{Yates {et~al.}(2020)Yates, Palmer, Manners, Boutle, Kohary, Mayne, \&
  Abraham}]{YatesEtAl2020}
Yates, J.~S., Palmer, P.~I., Manners, J., {et~al.} 2020, Monthly Notices of the
  Royal Astronomical Society, 492, 1691, \dodoi{10.1093/mnras/stz3520}

\bibitem[{Zhang {et~al.}(2017)Zhang, Tian, Wang, Dudhia, \&
  Chen}]{ZhangEtAl2017}
Zhang, X., Tian, F., Wang, Y., Dudhia, J., \& Chen, M. 2017, The Astrophysical
  Journal, 837, L27, \dodoi{10.3847/2041-8213/aa62fc}

\end{thebibliography}
\bibliographystyle{aasjournal}

\end{document}